%% file: main.tex
\documentclass[twocolumn]{aastex62} 

\usepackage{amsmath}	
\usepackage{enumitem}
\setlist[enumerate]{leftmargin=\parindent}
\setlist[itemize]{leftmargin=\parindent}

\newcommand{\Kepler}{\emph{Kepler}}
\newcommand{\TESS}{\emph{TESS}}

\newcommand{\Rsun}{\mbox{$R_{\odot}$}}

\renewcommand\_{\textunderscore\allowbreak}
\newcommand{\revision}{\color{blue}}

\begin{document}

\title[allesfitter]{\texttt{Allesfitter}: Flexible Star and Exoplanet Inference From Photometry and Radial Velocity}
\shorttitle{allesfitter}

\correspondingauthor{Maximilian N.\ G{\"u}nther}
\email{maxgue@mit.edu}

\author[0000-0002-3164-9086]{Maximilian N.\ G{\"u}nther}
\affil{Department of Physics, and Kavli Institute for Astrophysics and Space Research, Massachusetts Institute of Technology, Cambridge, MA 02139, USA}
\affil{Juan Carlos Torres Fellow}

\author[0000-0002-6939-9211]{Tansu Daylan}
\affil{Department of Physics, and Kavli Institute for Astrophysics and Space Research, Massachusetts Institute of Technology, Cambridge, MA 02139, USA}
\affil{Kavli Fellow}

\begin{abstract}

We present \texttt{allesfitter}, a public and open-source \texttt{python} software for flexible and robust inference of stars and exoplanets given photometric and radial velocity data. 
\texttt{Allesfitter} offers a rich selection of orbital and transit/eclipse models, accommodating multiple exoplanets, multi-star systems, transit-timing variations, phase curves, stellar variability, star spots, stellar flares, and various systematic noise models including Gaussian Processes.
It features both parameter estimation and Bayesian model selection,
allowing to easily run either a Markov Chain Monte Carlo (MCMC) or Nested Sampling fit.
For novice users, a graphical user interface allows to specify all input and perform analyses;
for \texttt{python} users, all modules can be readily imported in any existing scripts.
\texttt{Allesfitter} also produces publication-ready tables, LaTeX commands, and figures.
The software is publicly available (\url{https://github.com/MNGuenther/allesfitter}), \texttt{pip}-installable (\texttt{pip install allesfitter}) and well documented (\url{www.allesfitter.com}).
Finally, we demonstrate the software's capabilities on several examples and provide updates to the literature where possible for Pi~Mensae, TOI-216, WASP-18, KOI-1003 and GJ1243.

\end{abstract}

\keywords{planetary systems -- planets and satellites: general -- (stars:) binaries (including multiple) -- stars: flare -- Bayesian -- inference}

\section{Introduction}

With the wealth of available photometric and radial velocity (RV) observations from ground and space-based exoplanet missions, the analysis and modeling of data can become a limiting factor. Hence, the automation of this inference process in a reliable, scalable and reproducible way is crucial. The exoplanet community can especially profit from a user-friendly, all-in-one package that allows for fast and robust model comparison.

Various packages have been developed for forward-modeling of exoplanets and binary star systems, including \texttt{jktebop} \citep{Southworth2004a, Southworth2004b}, \texttt{pytransit} \citep{Parviainen2015}, \texttt{batman} \citep{Kreidberg2015}, \texttt{ellc} \citep{Maxted2016}, and \texttt{starry} \citep{Luger2019}. Using their underlying generative model of exoplanets and stars, these packages predict photometric and RV signals. General-purpose sampling algorithms can then be used to explore the parameter space of these forward-models consistent with observed data and subject to certain user-defined priors. 
The exoplanet community commonly uses Markov Chain Monte Carlo (MCMC) samplers (see Section~\ref{ss:MCMC}), with popular implementations including \texttt{emcee} \citep{Foreman-Mackey2013} and \texttt{pymc3} \citep{Salvatier2016}.
In this spirit, many researchers connect their forward-models with a sampler to analyze exoplanet-related data using private software (e.g. \texttt{mcmc} by \citealt{Gillon2012}, \texttt{gp-ebop} by \citealt{Gillen2017}, and \texttt{amelie} by \citealt{Hodzic2018}).
Only recently, the exoplanet community started to develop standardized public software, 
such as \texttt{exofast} \citep{Eastman2013, Eastman2019}, 
\texttt{allesfitter} (this work),
\texttt{juliet} \citep{Espinoza2019},
and \texttt{exoplanet} \citep{Foreman-Mackey2019}.
Despite their recency, all of these software have already been successfully and widely used in the literature.

Many existing software are focused on a specific task and for example applicable to exoplanet transits and RV signals, but often not to exoplanet phase curves, Rossiter-McLaughlin effects, brown dwarfs, low-mass binaries, star spots or stellar flares. 
Additionally, many software rely only on MCMC samplers. While MCMCs can be robust and fast, they generally do not provide statistically robust model comparison, given the absence of a low-variance estimate of the Bayesian evidence \citep[e.g.][]{Skilling2006}.

To overcome these caveats and provide general functionality and robustness, we developed the \texttt{allesfitter} package, which enables simultaneous (i.e., joint) inference of models for
\setlist{nolistsep}
\begin{itemize}[noitemsep]
    \item exoplanet transits, occultations, and RV signals,
    \item binary star eclipses and RV signals,
    \item transit-timing variations,
    \item phase curves,
    \item stellar variability,
    \item star spots,
    \item stellar flares,
    \item systematic noise, and
    \item injection-recovery tests,
\end{itemize}
and was already used in various publications \citep[e.g.][]{Huang2018b,Dragomir2019,Zhan2019,Guenther2019a,Guenther2019b,Shporer2019,Daylan2019,Badenas-Agusti2020}.

In a global analysis of both photometric and RV data, \texttt{allesfitter} also offers several ways to model red (systematic) noise, including polynomials, splines or Gaussian Processes (GPs).
Additionally, the software allows the user to choose between MCMC and various Nested Sampling algorithms. Both take fair samples from the posterior of the selected model, while the latter also  provides low-variance estimation of the Bayesian evidence for statistical model comparison and is more robust for high dimensionality (see Section~\ref{ss:Nested sampling} for details). 

Towards this purpose, \texttt{allesfitter} provides one framework uniting the versatile, publicly available packages 
\texttt{ellc} \citep[light curve and RV models;][]{Maxted2016}, 
\texttt{aflare} \citep[flare model;][]{Davenport2014},
\texttt{dynesty} \citep[static and dynamic Nested Sampling;][]{Speagle2020},
\texttt{emcee} \citep[MCMC sampling;][]{Foreman-Mackey2013} and 
\texttt{celerite} \citep[GP models;][]{Foreman-Mackey2017}.

A graphical user interface allows novice users to define all input parameters and settings without needing coding experience, making it well suited for undergraduate research programs, high school internships, or outreach events. However, users with \texttt{python} experience can import the package into their scripts and use its API. The outputs of \texttt{allesfitter} are publication-ready plots, ascii and LaTeX tables and LaTeX commands. The software is public and open source\footnote{GitHub: \url{https://github.com/MNGuenther/allesfitter}}, easily installable using \texttt{PyPi}\footnote{Installation: \texttt{pip install allesfitter}} and well documented\footnote{Documentation and tutorials: \url{www.allesfitter.com}}. Feedback and contributions are very welcome.

This paper is structured as follows. 
Section \ref{s:Bayesian statistics and sampling} introduces Bayesian statistics and the inference framework, while Section \ref{s:Generative model} discusses the forward models, including GPs. 
Section \ref{s:User Interface, Settings and Parameters} explains the user interface, underlying routine, parameters, and settings. 
Section \ref{s:Examples and case studies} showcases the application and performance for several test cases. In Section \ref{s:Summary and Conclusion} we discuss our results and conclude.

\section{Bayesian statistics and sampling}
\label{s:Bayesian statistics and sampling}

In Bayesian statistics, we compare models and infer their parameters using the `degree of belief' definition of probability \citep[see e.g.][]{MacKay2003}.
In this context, we are interested in the \textit{posterior probability} $\mathcal{P}(\theta) := P(\theta|M, D)$, i.e. the degree of belief about a set of parameters $\theta$ given a selected model $M$ and observed data $D$. The foundation for this inference problem is Bayes' theorem, which states that the posterior probability is given by
\begin{equation}
\underbrace{P(\theta|M, D)}_{\mathcal{P}} = 
\frac{ 
      \overbrace{P(D|M,\theta)}^{\mathcal{L}} 
      \overbrace{P(\theta| M)}^{\mathcal{\pi}} 
     }
     { 
      \underbrace{P(D|M)}_{\mathcal{Z}} 
     }.
\end{equation}
Here, the \textit{likelihood} $\mathcal{L}(\theta) := P(D|M,\theta)$ is the probability of observing the data $D$ under the given model $M$ with parameters $\theta$. 
The \textit{prior probability} $\mathcal{\pi}(\theta) := P(\theta| M)$ of the parameters $\theta$ given model $M$ encapsulates our knowledge of the model before the arrival of the data, $D$. Finally, $\mathcal{Z} := P(D|M)$ is the \textit{marginal likelihood}, also known as the \textit{Bayesian evidence}. It is calculated as the integral over the entire parameter space $\Omega_\theta$:
\begin{equation}
\underbrace{P(D|M)}_{\mathcal{Z}} = 
\int_{\Omega_\theta} \underbrace{P(D|M,\theta)}_{\mathcal{L}} 
     \underbrace{P(\theta| M)}_{\mathcal{\pi}} 
     \mathrm{d} \theta.
\end{equation}
It quantifies the degree of belief one should have about the model $M$ given the observed data $D$. Estimating the Bayesian evidence allows a comparison of different physical models. However, the integral is computationally expensive to solve.
While MCMC sampling completely bypasses its computation and leaves the Bayesian evidence unknown, Nested Sampling is specifically designed to estimate it (see below).

In the context of exoplanet science, the set of parameters $\theta$ may contain, for example, the orbital period, planet radius and stellar radius. The observed data $D$ may be time series such as the normalized light curve  and radial velocity of a target. The choice of priors can be motivated by other data sets or scaling arguments; for example, the period might be unknown but the stellar radius might be constrained by stellar models. 
Often, one would then assign a uniform prior to the period and a Gaussian prior on the stellar radius, with its mean and standard deviation reflecting the inference based on the characterization of the star using broad-band or high-resolution spectra.

The data-informed part of the posterior is the likelihood function. For $N$ data points $y_k \in (y_1, ..., y_N)$  with uncertainties $\sigma_k$ collected at times $t_k \in (t_1, ..., t_N)$, and a model evaluated on the same temporal grid $M(t)$, the logarithm of the likelihood is given as
\begin{equation}
\log{\mathcal{L}} = -\frac{1}{2}\sum_{k=1}^N \left[ \left( \frac{ y_k - M(t_k)}{\sigma_k} \right)^2 + \log{2\pi\sigma_k^2} \right].
\end{equation}
where we assume that uncertainties in data (light curves and RVs) have a Gaussian distribution.

\subsection{MCMC}
\label{ss:MCMC}

Markov Chain Monte Carlo (MCMC) methods are a class of tools for taking fair samples from a given probability distribution \citep[see e.g.][]{MacKay2003} by constructing a Markov chain, i.e., a memoryless sequence of elements ${\theta_0, \theta_1, ..., \theta_N}$. The statitonary distribution of this chain approximates the relevant probability distribution $P(\theta_{i+1}|\theta_i)$ in the limit $N\to\infty$. The memorylessness property requires that each new state $\theta_{i+1}$ depends only on the current state, $\theta_{i}$. Depending on the initial state of a Markov chain, the mixing of the chain will require a certain number of state transitions. But even after that, consecutive samples will have a non-vanishing autocorrelation. Therefore, after a Markov chain is constructed, the samples are split into a \textit{burn-in} and an \textit{evaluation} part, where the former samples are discarded since they are not draws from the posterior. The latter are used to estimate the posterior probability distribution only after thinning by a factor to ensure that the resulting chain is memoryless. The resulting chain yields the desired posterior that is used for parameter estimation. There are multiple ways to implement an MCMC algorithm. Examples include the Metropolis Hastins algorithm \citep{Metropolis1953, Hastings1970}, Gibbs sampling \citep{Geman1984}, and affine invariant sampling \citep{Goodman2010}. The common property of these algorithms is the concept of a random walk implemented via a proposal distribution in order to transition between such states. The proposal gets rejected with a probability set by how much it lowers the posterior and is accepted otherwise.

For \texttt{allesfitter}, we adopt the \texttt{emcee} package, which uses the \textit{affine invariant sampling} \citep[for details see][]{Goodman2010, Foreman-Mackey2013}. This enables efficient sampling from potentially skewed posterior probability distributions with correlated parameters and precludes the necessity to specify a proposal scale for each parameter. To do so, it employs multiple walkers (i.e., chains) with leap-frog proposals to explore the parameter space. The default settings for \texttt{allesfitter}'s MCMC implementation can be found in Table~\ref{tab:Settings}.

\subsubsection{Assessing convergence}
\label{sss:Assessing convergence MCMC}

Despite discarding the initial samples and thinning the remaining chain, the resulting chain may still not be fully mixed \citep[see][]{Goodman2010, Foreman-Mackey2013}. Therefore, confirming their independence and thus the convergence of the MCMC sampling is important, yet often not strictly mathematically possible. To assess convergence nevertheless, two commonly used approximate criteria are requiring a maximum autocorrelation or Gelman Rubin test statistic. 
In \texttt{allesfitter}, we implement the integrated autocorrelation time as the convergence criterion. Using it, we estimate the effective number of independent samples in the chain. It is recommended that the user runs the MCMC chains until all parameters have a chain length of at least 30 times the autocorrelation time \citep[see][]{Foreman-Mackey2013}.

\subsubsection{Limitations for model selection}
\label{sss:Limitations for model selection}
MCMC is an efficient tool to take samples from the posterior of a model given some data. In addition to parameter estimation, one might also want to compare two models, e.g., a model of RV data with one and two planets, respectively. However, MCMC is limited when estimating the overall degree of belief in the associated model when using the harmonic mean, because the contribution of rare samples from the posterior make increasingly large contributions to the Bayesian evidence \citep[see e.g.][]{Weinberg2010}. This makes it hard to compare different models by estimating the \emph{Bayes factor}, i.e., the ratio of the Bayesian evidence of the models.

\subsection{Nested sampling}
\label{ss:Nested sampling}

Nested Sampling is an inference algorithm to directly estimate the Bayesian evidence by sampling from the prior subject to evolving constraints on likelihood \citep{Skilling2004, Skilling2006, Feroz2009, Feroz2019, Handley2015, Higson2018, Higson2019}. Its low-variance estimate of the Bayesian evidence allows it to be used for robust model comparison. In the exoplanet context, this then enables model tests such as comparing models with different numbers of exoplanets \citep[see][]{Hall2018}, a circular orbit against an eccentric orbit, or transit times with TTVs to those without TTVs. 

Nested Sampling achieves this by avoiding to sample directly from the posterior. Instead, it divides the problem into a series of simpler sampling problems. First, it draws a number of \textit{live points} from the prior. Next, the live point with the lowest likelihood is removed. A new live point is created by sampling from the prior while requiring that it has a higher likelihood than before. The algorithm iterates over this process until the change in the resulting Bayesian evidence is below a certain threshold. The resulting samples from \textit{nested slices} are sorted according to their likelihoods and used to compute the evidence integral by rewriting the multi-dimensional marginalization integral as a one-dimensional integral over the \textit{prior mass} $X$ of the hypervolume defined by points with a likelihood larger than likelihood threshold of each slice,
\begin{equation}
    Z = \int_{\Omega_\theta} \mathcal{L}(\theta) \mathcal{\pi}(\theta) d\theta = \int_0^1 L(X) dX.
\end{equation}
Here, the prior volume $X$ is defined as the fraction of the prior where the likelihood $L(\theta)$ is greater or equal to a certain threshold $\lambda$, i.e., 
\begin{equation}
    X(\lambda) = \int_{\Omega_\theta: \mathcal{L}(\Omega)>\lambda} \mathcal{\pi}(\theta) d\theta,
\end{equation}
where $\mathcal{L}(X)$ is the iso-likelihood. The bounds of the integral are defined by the normalization of the prior, leading to $X(\lambda=0)=1$ and $X(\lambda \rightarrow \infty)=0$. Nested Sampling then uses a statistical approach to generate samples from the prior $\mathcal{\pi}(\theta)$. With this, it can approximate the prior volume $X(\theta)$ and its differential, and thus to compute the evidence integral.

\texttt{allesfitter} implements the \texttt{dynesty} package, which offers both static and dynamic Nested Sampling, as well as multiple options such as slicing, multinest or polynest algorithms \citep[for details see][]{Speagle2020}. Dynamic Nested Sampling, in particular, is recommended as a generalisation of Nested Sampling in which samples can be drawn more efficiently by varying the number of live points. The default settings for \texttt{allesfitter}'s Nested Sampling implementation can be found in Table~\ref{tab:Settings}.

\subsubsection{Assessing convergence}
\label{sss:Assessing convergence NS}
In Nested Sampling, the algorithm is considered converged once the gain in logarithmic Bayesian evidence, $\Delta \ln{\mathcal{Z}}$, is below a certain threshold. For \texttt{allesfitter}, we recommend the default threshold of $\Delta \ln{\mathcal{Z}} \leq 0.01$ \citep[following \texttt{dynesty};][]{Speagle2020}.

\subsubsection{Model selection}
\label{sss:Model selection}
Because Bayesian evidence $\mathcal{Z} := P(D|M)$ is marginalized over the parameters of a given model, it allows us to compare models given the same data. This can be done by calculating the Bayes factor $\mathcal{R}$, which is defined as 
\begin{equation}
    R := \frac{ \mathcal{Z}_\mathrm{Model\,1} }{ \mathcal{Z}_\mathrm{Model\,2} }
         \frac{ \mathcal{\pi}_\mathrm{Model\,1} }{ \mathcal{\pi}_\mathrm{Model\,2} },
\end{equation}
where $\mathcal{Z}_\mathrm{Model\,1}$ and $\mathcal{Z}_\mathrm{Model\,2}$ are the Bayesian evidence for each model (.e.g., a one-planet versus a two-planet model), and $\mathcal{\pi}_\mathrm{Model\,1}$ and $\mathcal{\pi}_\mathrm{Model\,1}$ are the prior beliefs in each model (not to be confused with the prior density for a set of parameters of a model).

\citet{Jeffreys1998} and \citet{Kass1995} suggest that, given a null model $M_1$, the alternative (more complex) model $M_2$ should only be selected if there is sufficient relative Bayesian evidence for it as quantified by $\ln{R} \gtrsim 5$. In detail, they suggest the interpretation given in Table~\ref{tab:Bayes_factors}.

\begin{table}[ht]
    \centering
	\caption{Interpretation of Bayes' factors from \citet{Jeffreys1998} and \citet{Kass1995}.}
	\centering
	\begin{tabular}{ll}
    \hline
    \hline
    $\ln{R}$ & Strength of evidence\\
    \hline
    0 to 1.2 & Barely worth mentioning \\
    1.2 to 2.3 & Substantial\\
    2.3 to 4.6 & Strong to very strong\\
    $>$\,4.6 & Decisive \\
    \hline
	\end{tabular}
   \label{tab:Bayes_factors}
\end{table}

\section{Generative models}
\label{s:Generative model}

\subsection{Orbits, eclipses/transits/occultations and stellar brightness features}

Our forward model for the photometry and radial velocity observed in stellar and planetary systems is largely implemented via the public, open-source software \texttt{ellc} \citep{Maxted2016}. At its core, \texttt{ellc} is a fast, flexible and accurate binary star model, which is also readily applicable to exoplanet models (in the limit of the companion being small, low-mass, and faint). The software implements a transit model as well as incorporating the effects of star spots, Doppler boosting, light-travel times, the flux-weighted radial velocity during an eclipse (Rossiter-McLaghlin effect) and light from a blended source. The \texttt{ellc} generative model was substantially tested, compared with other existing models, and already widely used in the literature. We summarize the core principles of \texttt{ellc} in the following, and refer the reader to \citet{Maxted2016} for all details.

\texttt{ellc} models the stars as triaxial ellipsoids, and calculates their flux using Gauss-Legendre integration over the visible surface. The shape of the objects can be calculated in three ways:

\begin{itemize}
    \item In the spherical limit,
    \item Using the Roche potential, including non-synchronous rotation,
    \item a polytropic equation of state. 
\end{itemize}

In \texttt{allesfitter}, this flexibility allows the user to adjust the settings to their desired methods, and to readily model, for example, the ellipsoidal modulation in the phase curve of a binary star or hot Jupiter system.
To compute the object positions, \texttt{ellc} follows Keplerian orbits with fixed orbital eccentricity $e$ and an apsidal motion when provided. The positions are updated using Kepler’s equation, $M = E - e \sin{E}$, in order to solve for the eccentric anomaly $E$ from the mean anomaly $M = 2 \pi (t_i - t_0)/P_a$ for a fixed anomalistic period $P_a$, and to compute the true anomaly $\nu$. This approach incorporates the calculation of, and a correction for, the light travel times, which enables us to compute Doppler boosting effects.
Furthermore, the surface brightness distribution $I_\lambda(s,t)$ at any surface point $s$ at time $t$ incorporates established limb darkening laws $U(\mu)$, gravity darkening $G(s,t)$, and the irradiation of the body by its companion $H(s,t)U_H(\mu)$ following the relation
\begin{equation}
    I_\lambda(s,t) = I_0 U(\mu) G(s,t) + H(s,t) U_H(\mu).
\end{equation}
The limb darkening laws include all standard choices, from constant to four-parameter laws, depending on the normalized distance $\mu$ from the center (see Table~\ref{tab:Limb_darkening_laws}). Moreover the gravity darkening calculation assumes that the specific intensity relates to the local gravity by a wavelength-dependent power law.
For this, \texttt{ellc} calculates the local gravity via the gradient of the Roche potential. The user decides whether the local gravity should be done sampled for all grid points (computationally expensive), or via the interpolation of a few samples on the stellar surface (default). Computing the irradiation of a body by its companion comes with many caveats, as the incident energy can change the thermal structure of the atmosphere and the emergent spectrum can differ substantially from the incident radiation. To solve this, \texttt{ellc} simplifies this problem with a parameterization that relates directly to the specific intensity distribution and angular orientation \citep[for details see][]{Maxted2016}.  

\begin{table}[ht]
    \centering
	\caption{Common limb darkening laws.}
    \label{tab:Limb_darkening_laws}
	\centering
	\begin{tabular}{ll}
    \hline
    \hline
    Name & Equation for $U(\mu)$\\
    \hline
    constant & $1$\\
    linear$^1$ & $1 - c_1(1 - \mu)$\\
    square-root$^2$ & $1 - c_1(1 - \mu) - c_2/(1 - \sqrt{\mu})$ \\
    exponential$^3$ & $1 - c_1(1 - \mu) - c_2/(1 - e^\mu)$ \\
    logarithmic$^4$ & $1 - c_1(1 - \mu) - c_2\mu \ln{\mu}$ \\
    quadratic$^5$ & $1 - c_1 (1 - \mu) - c_2 (1 - \mu)^2$ \\
    three-parameter$^6$ & $1 - c_1 (1 - \mu) - c_2(1 - \mu)^{3/2}$ \\
                        &$- c_3(1 - \mu)^2$ \\
    four-parameter$^7$ & $1 - c_1(1 - \mu)^{1/2} - c_2 (1 - \mu)$ \\
                       &$- c_3(1 - \mu)^{3/2} - c_4(1 - \mu)^2$ \\
    \hline
	\end{tabular}
    \begin{minipage}{4.3in} 
    $^1$\citet{Schwarzschild1906}\\
    $^2$\citet{Diaz-Cordoves1992}\\
    $^3$\citet{Claret2003}\\
    $^4$\citet{Klinglesmith1970}\\
    $^5$\citet{Kopal1950}
    $^6$\citet{Sing2010}
    $^7$\citet{Claret2000}
    \end{minipage}
\end{table}

Doppler boosting (relativistic beaming) and Doppler shift are relativistic effects which increase and decrease, respectively, the observed flux depending on the bodies' radial velocity, spectra and observation wavelengths.
By explicitly modelling all light travel times in the system, \texttt{ellc} thus enables us to compute these effects for photometric phase curves and the Rossiter-McLaughlin effect in the radial velocity. Star spots are approximated as circular regions with different brightness, set at a given longitude and latitude on the star. 
Similar to the bodies' shapes, the spot brightness is integrated over triaxial ellipsoids taking into account all limb darkening requirements. Finally, light curves are generated by integrating over the bodies' surface brightness distributions using a mix of Gaussian-Legendre and analytical integration methods. This calculation incorporates the dilution by a third light term, originating from another body in the system. 

Additionally, RV signals are generated by assuming Keplerian orbits of the bodies. The user can choose whether these are weighted by the flux from the visible surface (default). These computations incorporate terms for the projected rotational velocity (both equatorial and asynchronous), different shapes of the bodies, and star spots on their surfaces.

\subsection{Stellar flares}

Stellar flares are explosive magnetic reconnection events that emit large amounts of radiation, predominately in the UV to X-ray spectrum. They are the product of the stellar magnetic dynamo, driven by the sheering rotation of the star's radiative core and convective envelope. Flares on M dwarf stars are especially much more frequent and energetic than on our Sun, posing disadvantages as well as opportunities, for the genesis and survival of life on exoplanets \citep[e.g.][]{Pettersen1989,Rimmer2018,Guenther2019a}.

To model stellar flares, we adopt the \texttt{aflare} module from the \texttt{appaloosa} package \citep{Davenport2014, Davenport2016}. This model only depends on three parameters: the flare's peak time $t_\mathrm{peak}$, amplitude $A$ and full-width at half-maximum $t_{1/2}$. This empirical template was created from a sample of 885 flares on the M dwarf GJ 1243, assuming that stellar flares share a common formation mechanism across stellar types. The authors selected `classical flares' in the Kepler light curve with a duration between 20 and 75 minutes. They then detrended the data to remove modulation by star spots and normalized all flares to scale from 0 to 1 in amplitude, and by a single time scale factor in width to fulfill the normalization $t_{1/2}=1$.
Next, they fitted a third order polynomial to describe the rise time as
\begin{equation}
    F_\mathrm{rise} = 1 + c_1 t_{1/2} + c_2 t_{1/2}^2 + c_3 t_{1/2}^3 + c_4 t_{1/2}^4,
\end{equation}
and a double exponential function to describe the decay time as
\begin{equation}
    F_\mathrm{decay} = 1 + c_1 e^{- c_2 t_{1/2}} + c_3 e^{- c_4 t_{1/2}}.
\end{equation}

The rapid rise towards the peak flux was motivated by the morphology seen in ground-based white-light photometry \citep[e.g.][]{Kowalski2013}. We note that this does not hold true for all flares, as for example, pointed out by \citep{Jackman2018, Jackman2019}, who used high-cadence photometry and found the need for additional terms to describe a `roll over' rather than a sharp peak. The double-exponential decay represents two physically distinct regions with independent exponential cooling profiles. \citet{Davenport2014} argue that the initial decay might be dominated by a hotter region which cools rapidly, and that the gradual decay stems from a cooler region which cools slower.

\subsection{Red noise, stellar variability and Gaussian Processes (GP)}
\label{ss:Red noise, stellar variability and Gaussian Processes (GP)}

Generative models used to fit observations are never perfect descriptions of the data, even up to \emph{white} (uncorrelated) noise, because observed data are always affected by physical processes not available in the fitted model. The total effect of these unaccounted processes in the data is usually referred to as \emph{red} (correlated) noise. The apparent correlation of this noise is a consequence of the time-variability of the unaccounted physical processes. Examples include instrumental noise, atmospheric scintillation effects for ground-based observatories, light from blended objects, or scattered light from the Earth and the moon for space-based observatories. Stellar variability may also be counted as correlated noise if one is only interested in the properties of an exoplanet but not the star's behavior. However, at other times, one may wish to model stellar variability explicitly to characterize the stellar rotation or activity.

\texttt{Allesfitter} includes various options to model red noise and stellar variability, including constant offsets, polynomial trends, cubic splines, and various GP models. An overview of all these models is given in Table~\ref{tab:Red_noise_and_stellar_variability}. In the software interface, these models are implemented in two complementary ways:
\begin{enumerate}
\item a baseline model, which is instrument dependent (see Section~\ref{ss:Baselines (red noise)}) and
\item a stellar variability model, which fits a common trend across all instruments (see Section~\ref{ss:Stellar variability}).
\end{enumerate}

\begin{table}[ht]
    \centering
	\caption{Red noise and stellar variability models $M(t)$ in dependency of time $t$. Other variables are the fit parameters explained in the text.}
    \label{tab:Red_noise_and_stellar_variability}
	\centering
	\begin{tabular}{ll}
    \hline
    \hline
    Name & Equation for $M(t)$\\
    \hline
    none & $0$\\
    offset & $c_1$\\
    linear & $c_1 + c_2 t$\\
    quadratic & $c_1 + c_2 t + c_3 t^2$\\
    third-order poly. & $c_1 + c_2 t + c_3 t^2 + c_4 t^3$\\
    fourth-order poly. & $c_1 + c_2 t + c_3 t^2 + c_4 t^3 + c_5 t^4$\\
    cubic spline & $c_1(t) + c_2(t) t + c_3(t) t^2 + c_4(t) t^3$\\
    GP real & $a e^{-c t}$\\
    GP complex & $\frac{1}{2}\left[\left(a+b\right) e^{-\left(c+d\right) t} \right.$ \\
               & ~~~ $+ \left. \left(a-b\right) e^{-\left(c-d\right) t}\right]$\\
    GP Mat{\'e}rn 3/2 & $\sigma^{2} \left[ (1+1 / \epsilon) e^{-(1-\epsilon) \sqrt{3} t / \rho} \right.$\\
                  & ~~~ $\left. (1-1 / \epsilon) e^{-(1+\epsilon) \sqrt{3} t / \rho} \right]$, $\epsilon=0.01$\\
    GP SHO & $\sqrt{\frac{2}{\pi}} \frac{S_{0} \omega_{0}^{4}}{\left(t^{2}-\omega_{0}^{2}\right)^{2}+\omega_{0}^{2} t^{2} / Q^{2}}$\\
    
    \hline
	\end{tabular}
\end{table}

Some of the most versatile baseline or stellar variability models are GPs \citep{Rasmussen2005, Bishop2006, Gibson2012, Roberts2013}.
Instead of fitting for the parameters of a chosen model (e.g. coefficients of a polynomial), GP regression fits for a family of functions to marginalize over the choice of the basis in a so-called non-parametric approach.

In Bayesian inference, a GP can be interpreted as a prior on the space of functions that describe the data \citep[see e.g.][]{Murphy2012}. When updated based on observed data, the posterior of the GP model characterizes the distribution of baselines needed to fit the data along with the desired physical model. The autocorrelation of the GP is described by the specified distance metric and kernel, i.e., a covariance matrix which sets the flexibility of the GP. Kernels with a large autocorrelation scale produce smoother baselines, whereas those with a small autocorrelation scale produce turbulent baselines. The GP kernel is fitted to the data by sampling from the posterior of the hyperparameters. This posterior can then be linked to physical processes such as stellar variability, atmospheric scintillation or instrumental noise.

\texttt{Allesfitter} implements the \texttt{celerite} package, which provides a series expressions of typical GP kernels, achieving a significant improvement in execution time \citep[see][for details]{Foreman-Mackey2017}. Out of the available kernel functions (see Table~\ref{tab:Red_noise_and_stellar_variability} for equations), the \textit{real} kernel is the simplest, with exponential decay and two hyper-parameters, the amplitude $a$ and time scale $c$. The \textit{complex} kernel is a relatively more complex model with amplitudes $a$ and $b$, and time scales $c$ and $d$. The \textit{Mat{\'e}rn 3/2} kernel is one of the most versatile and frequently used kernel in astronomy, as it can describe smooth long-term trends as well as stochastic short-term variations. It features two hyper-parameters, i.e., the amplitude scale $\sigma$ and length scale $\rho$.
Lastly, the \textit{Simple Harmonic Oscillating (SHO)} kernel is appropriate for (semi-)periodic signals, as it represents a stochastically-driven, damped harmonic oscillator. Its hyper-parameters are the amplitude scale $S_0$, the damping $Q$ and the frequency $\omega$.

\section{User Interface, Settings and Parameters}
\label{s:User Interface, Settings and Parameters}


\subsection{Overview}

Every \texttt{allesfitter} run is designed to operate in a user-designated working directory. The input configuration to \texttt{allesfitter} is provided via comma separated value (CSV) files for the settings and parameters in this working directory, named \textit{settings.csv} and \textit{params.csv} respectively. All possible inputs in these files are explained in Tables~\ref{tab:Settings} and \ref{tab:Parameters}, and any special implementations are laid out in the sections below.

All data must be stored as CSV files in this working directory. The data file names must match those provided in the settings and parameters files. For example, if the user names the instruments \TESS{} and ESPRESSO in those files, the data file names must be \textit{TESS.csv} or \textit{ESPRESSO.csv}. 
Light curve files need three columns: the time (in days), relative flux (i.e., normalized to 1), and the uncertainty of the relative flux. The errors are only needed for their relative values across time since the errors are scaled (mean of all errors) using a model parameter (see Section~\ref{ss:White noise and jitter terms} and Table~\ref{tab:Parameters}). Therefore, if the errors are unknown, the final column can be filled with values of 1.
Radial velocity files also need three columns: the time (in days), radial velocity (in km/s), and instrumental error of the radial velocity. For radial velocity instruments, both the error weights and their scaling will affect the fits. If the instrumental noise is unknown, the final column can be filled with values of 0, as a stellar jitter term will still be added in quadrature during the fit (see Section~\ref{ss:White noise and jitter terms} and Table~\ref{tab:Parameters}).

There are two ways to start an \texttt{allesfitter} run. First, the graphical user interface (GUI) can guide the user through the entire setup, from assigning a working directory to generating the necessary settings and parameters files to running the analyses. Second, the user can manually create the settings and parameter files in the working directory, either from scratch or by using any of the template files. Then the user can use the application programming interface (API) to import the \texttt{allesfitter} module and call all respective functions to start the run (see below). When a run is started, \texttt{allesfitter} creates a \textit{results} folder in the base folder. A log file is created for each run, uniquely named with the ISO 8601 compliant date and time. All output will also be saved into this folder (see below).

\subsection{API and Output}

\texttt{Allesfitter} is built in a modular way, and as such, many functions can be called directly from the \texttt{python} interface. In this section, we briefly lay out the most important parts of the Application Programming Interface (API), and refer the user to \url{www.allesfitter.com} for details, future updates and the most up-to-date documentation.

As a first step after setting up the working directory, the users can investigate how well their initial guess matches the data by calling
\begin{equation*}
    \textit{allesfitter.show\_initial\_guess(datadir),}
\end{equation*}
where \textit{datadir} is the working directory path. This creates initial guess plots and a logfile in the working directory. Once the user verifies that the data and initial guess look as intended, the inference (MCMC or Nested Sampling) can be initiated.

An MCMC fit can be performed by calling
\begin{equation*}
    \textit{allesfitter.mcmc\_fit(datadir).}
\end{equation*}
This creates the file \textit{mcmc\_save.h5} and a logfile in the given directory path, and the state of the sampling can be monitored with a waitbar in the \textit{python} console. Any time during the execution, output files can be created by calling
\begin{equation*}
    \textit{allesfitter.mcmc\_output(datadir).}
\end{equation*}
including the samples up to the last stored state. This allows an efficient way to diagnose whether the run is configured and behaves as intended, e.g., by inspecting the evolution of the chains. Once the sampling is completed, this should be re-executed to generate the final results.

A Nested Sampling fit can be performed by calling
\begin{equation*}
    \textit{allesfitter.ns\_fit(datadir).}
\end{equation*}
This creates a logfile in the given directory path, and the state of the sampling can be monitored through the sampling output in the \textit{python} console. Once the sampler has converged, the samples will be stored in the file.
Note that, due to Nested Sampling's iterative algorithm conditional on convergence, the progress can not be monitored in a waitbar. However, the progress can be gauged by monitoring how the value \textit{dlogZ} decreases down to the chosen tolerance limit (default: 0.01). As a rule of thumb, the time needed for completion scales logarithmically, such that, for example, decreasing from 100 to 10 takes the same time as from 10 to 1. Once the algorithm converges, all output files can then be created by calling
\begin{equation*}
    \textit{allesfitter.ns\_output(datadir).}
\end{equation*}

A helpful feature for fine-tuning figures and retrieving results from converged runs is the \textit{allesclass}, which can be called as
\begin{equation*}
    \textit{allesfitter.allesclass(datadir).}
\end{equation*}
This allows the user to easily retrieve all data, parameters, settings, initial guess and posterior samples, as well as the baseline, stellar variability, transit and phase curve forward-model samples. It also offers various plotting utilities to easily customize figures for publications.

Other major modules of \texttt{allesfitter} are the easy-to-use interfaces for transit injection with \texttt{ellc} and recovery with \texttt{tls}. We refer the user to \url{www.allesfitter.com} for detailed documentation on these modules.
Additionally, \texttt{allesfitter} contains various modules to process light curve and RV data, and perform tasks such as transit masking and phase folding.

\subsection{Transit/eclipse epoch}
\label{ss:Transit/eclipse epoch}
In a linear ephemeris model, the transit times are described by an epoch and a period. Shifting the transit epoch into the middle of the temporal interval of the data reduces the degeneracy between the epoch and period. However, the epoch is often reported as the time of the first transit in the literature or archives. Therefore, \texttt{Allesfitter} automatically shifts the epoch into the middle of the data set if the user sets \textit{shift\_epoch} to \textit{True} in the settings file (Table~\ref{tab:Settings}).

\subsection{Limb darkening parameterizations}
\label{ss:Limb darkening parameterizations}
\texttt{allesfitter} allows to chose either a constant, linear, quadratic or three-parameter limb darkening law (Tables~\ref{tab:Settings} and \ref{tab:Parameters}). It takes as input the transformed limb darkening coefficients $(q_1, q_2, q_3)$, which refer to the parameterization from \citet{Kipping2013} and \citet{Kipping2017}.
We recommended the users to sample $(q_1, q_2, q_3)$ with uniform priors between $[0,1]$, and let the data inform the limb darkening model parameters. Alternatively, a user might wish to rely on tabulated values for $(u_1, u_2, u_3)$ \citep[e.g.][]{Claret2013}. If so, the user has to first transform these values and their priors into $(q_1, q_2, q_3)$ before passing them as inputs into \texttt{allesfitter}. For a linear law \citep{Schwarzschild1906}, the transformation is  $q_1 = u_1$, whereas for a quadratic law \citep{Kopal1950}, the transformation between ($u_1$,$u_2$) and ($q_1$,$q_2$) uses the following equations \citep{Kipping2013}: 
\begin{equation}
\begin{aligned}[c]
    u_1 &= 2 \sqrt{q_1} q_2\\
    u_2 &= \sqrt{q_1} (1 - 2 q_2)
\end{aligned}
\quad\Leftrightarrow\quad
\begin{aligned}[c]
    q_{1} &= \left(u_{1}+u_{2}\right)^{2}\\
    q_{2} &= 0.5 u_{1}\left(u_{1}+u_{2}\right)^{-1}
\end{aligned}
\end{equation}
For the three-parameter law \citep{Sing2010}, transformation algorithms are provided by \citet{Kipping2017}.

After convergence, \texttt{allesfitter} recomputes the physical parameters ($u_1,u_2,u_3$) for comparison and interpretability (see Section~\ref{ss:Derived parameters}).

\subsection{White noise and jitter terms}
\label{ss:White noise and jitter terms}
The photometric uncertainties the user inputs are normalized to 1, such that only their weights towards another are important. The mean of the uncertainties are fitted as a model parameter, 
\begin{equation}
    \underbrace{\vec{ \sigma }_\mathrm{white, total}}_{\mathrm{result}}  
    = 
    \underbrace{\vec{ \sigma }_\mathrm{white, weights}}_{\mathrm{user~input}}
    \cdot
    \underbrace{\sigma_\mathrm{white, scaling}}_{\mathrm{fit~param.}} 
\end{equation}

In contrast, for RV data, a jitter term is fitted. Therefore, the input values are not normalized. Instead, the total uncertainty on each RV data point is calculated as 
\begin{equation}
    \underbrace{\vec{ \sigma }_\mathrm{white, total}}_{\mathrm{result}}  
    = 
    \sqrt{ 
    \underbrace{\sigma_\mathrm{white, inst}^2}_{\mathrm{user~input}} 
    + 
    \underbrace{\sigma_\mathrm{jitter}^2}_{\mathrm{fit~param.}} 
    } 
\end{equation}

\subsection{Baselines (red noise)}
\label{ss:Baselines (red noise)}
Various baseline models are available to handle red noise caused by instrumental systematics and stellar variability. While these models are described in detail in Section~\ref{ss:Red noise, stellar variability and Gaussian Processes (GP)}, we here explain how they can be called via the API. In the settings file, the user can choose between options from two major groups (see Table~\ref{tab:Settings}):
\begin{itemize}
    \item sampling from the posterior of the parameters that describe the baseline (called \textit{sample\_*}),
    \item profiling the likelihood by maximizing it for each proposal (called \textit{hybrid\_*}).
\end{itemize}

For all \textit{sample\_*} options, the user must also provide the respective parameters in the parameter file (see Table~\ref{tab:Parameters}). All available baseline options are:
\begin{itemize}

    \item No baseline fitting. The respective setting is \textit{none}, and the baseline is fixed at 1 for light curve data, and at 0 for RV data.
    
    \item Sampling a constant offset. The respective setting is \textit{sample\_offset} and the corresponding parameter is \textit{baseline\_offset\_[key]\_[inst]}.
    
    \item Sampling a linear trend. The respective setting is \textit{sample\_linear} and the two corresponding parameters are \textit{baseline\_offset\_[key]\_[inst]} and \textit{baseline\_slope\_[key]\_[inst]}.
    
    \item Sampling a GP with a real kernel. The respective setting is \textit{sample\_GP\_real} and the two corresponding parameters are
    \textit{baseline\_gp\_real\_lna\_[key]\_[inst]} and
    \textit{baseline\_gp\_real\_lnc\_[key]\_[inst]}.
    
    \item Sampling a GP with a complex kernel. The respective setting is \textit{sample\_GP\_complex} and the four corresponding parameters are 
    \textit{baseline\_gp\_complex\_lna\_[key]\_[inst]},
    \textit{baseline\_gp\_complex\_lnb\_[key]\_[inst]},
    \textit{baseline\_gp\_complex\_lnc\_[key]\_[inst]} and
    \textit{baseline\_gp\_complex\_lnd\_[key]\_[inst]}.
    
    \item Sampling a GP with a Mat{\'e}rn-3/2 kernel. The respective setting is \textit{sample\_GP\_Matern32} and the two corresponding parameters are \textit{baseline\_gp\_matern32\_ lnsigma\_[key]\_[inst]} and \textit{baseline\_gp\_matern32\_lnrho\_[key]\_[inst]}.
    
    \item Sampling a GP with a simple harmonic oscillator (SHO) kernel. The respective setting is \textit{sample\_GP\_SHO} and the two corresponding parameters are
    \textit{baseline\_gp\_sho\_lnS0\_[key]\_[inst]},
    \textit{baseline\_gp\_sho\_lnQ\_[key]\_[inst]} and
    \textit{baseline\_gp\_sho\_lnomega0\_[key]\_[inst]}.
    
    \item Hybrid offset. The respective setting is \textit{hybrid\_offset}. At each step, the median of the residuals will be taken as the baseline.
    
    \item Hybrid polynomials. The respective setting is \textit{hybrid\_poly\_*} followed by a number from 1 to 4, which sets the degree of the polynomial. At each step, a least squares minimization will determine the polynomial parameters to set the baseline.
    
    \item Hybrid cubic spline. The respective setting is \textit{hybrid\_spline}. At each step, a least squares minimization will determine the cubic spline parameters to set the baseline.
    
\end{itemize}

\subsection{Stellar variability}
\label{ss:Stellar variability}
Stellar variability can generate a signal or red noise that is shared between different instruments, especially for those in similar bands. Hence, it is implemented as separate component in addition to the baselines for individual instrument. For example, the user may wish to fit two data sets from different instruments that have distinct instrumental red noise, but a common stellar variability trend.

The functionality is the same as for baselines (see Section~\ref{ss:Baselines (red noise)} and Tables~\ref{tab:Settings}\&\ref{tab:Parameters}). The user only needs to replace they keyword \textit{baseline} with \textit{stellar\_var} and drop the part \textit{\_[inst]}. For example, for a GP with Mat{\'e}rn 3/2 kernel, the setting is \textit{sample\_GP\_Matern32} and the two corresponding parameters are \textit{stellar\_var\_gp\_matern32\_ lnsigma\_[key]} and \textit{stellar\_var\_gp\_matern32\_lnrho\_[key]}.

\subsection{External priors: stellar host density}
\label{ss:External priors: stellar host density}

If enabled by the user, an external normal prior on the host's bulk density is calculated from the input stellar radius and mass (by setting \textit{use\_stellar\_density\_prior} to \textit{True} and passing a stellar parameters file; see Table~\ref{tab:Settings}). 
At each proposal and for each companion, this is compared to the host's bulk density $\rho_\star$ derived via \citet{Seager2003} as:
\begin{equation}
\rho_\star = 3 \pi \left(\frac{a}{R_\star}\right)^3 P^{-2}.
\end{equation}
Here, $R_\star$ is the host's radius and $a$ and $P$ are the semi-major axis and orbital period for this companion, which has a radius $R_\mathrm{comp}$. Since this is only valid for $\left(\frac{R_\mathrm{comp}}{R_\star}\right)^3 \rightarrow 0$, \texttt{allesfitter} only applies this external prior if $\frac{R_\mathrm{comp}}{R_\star} \lesssim 0.22$, allowing a $<1$\% error.

\subsection{Phase curves}
\label{ss:Phase curves}

\texttt{allesfitter} offers three options for modeling exoplanet and binary star phase curves. 
First, a parametric method can be used to fit a linear combination of sine and cosine waveforms. The semi-amplitudes of these terms can then be interpreted as physical quantities, which is a common approach in exoplanet phase curve analyses (Section~\ref{sss:Phase curves using sines}).
Second, a similar but transformed sinusoidal parametrization can be chosen to ensure that the user input directly relates to physical quantities (Section~\ref{sss:Phase curves using transformed sines}).
Third, a physical model can be employed by utilizing the forward-model of \texttt{ellc} (Section~\ref{sss:Phase curves with ellc's physical model}).

\subsubsection{Phase curves using sines}
\label{sss:Phase curves using sines}

One can approximate a phase curve as a linear combination of sine and cosine waveforms, which models the out-of-eclipse variation of the system's flux, $F$, as a third-order harmonic series dependent on the orbital phase $\phi (t)$ \citep[e.g.][]{Carter2011, Shporer2019, Wong2020}:
\begin{align}
    F \propto \sum_{k=1}^{3} A_{k} \sin{k \phi(t)} + \sum_{k=1}^{3} B_{k} \cos{k \phi(t)} 
\end{align}

These terms can be related to the following three physical effects:
\begin{itemize}
    \item Doppler boosting (beaming) modulation, which is caused by the periodic redward and blueward color shifts of the emission from the host star due to its orbital motion \citep[e.g][]{Shakura1987}. The effect can be approximated by the sinusoidal term $A_1 \sin{(\phi(t))}$.
    Only positive values of $A_1$ allow a physical interpretation as the semi-amplitude of the host star's beaming effect, $A_\mathrm{beam}^\mathrm{semi} = A_1$.
    
    \item Atmospheric modulation, which includes the thermal and reflected emission from the companion \citep[e.g.][]{Snellen2009}. It can be approximated by the fundamental cosine term $B_1 \cos{(\phi(t))}$, where $B_1$ is a semi-amplitude.
    Only negative values of $B_1$ allow physical interpretation as the full (peak-to-peak) amplitude of the companion's atmospheric component, $A_\mathrm{atmo}^\mathrm{full} = -2 B_1$.
    
    \item Ellipsoidal modulation, which appears when the host star is tidally distorted due to the gravity of the companion \citep[e.g.][]{Morris1985}. It can be approximated by the sum of harmonic cosine terms, with the leading-order term being $B_2 \cos{(2 \phi(t))}$ and the next-order term being $B_3 \cos{(3 \phi(t))}$. Note that the leading-order term is sufficient for exoplanet phase curves, but the next-order term can become detectable for binary phase curves.
    Only negative values of $B_2$ and $B_3$ allow physical interpretation as the system's ellipsoidal components, $A_\mathrm{elli;1st}^\mathrm{full} = -2 B_2$, $A_\mathrm{elli;2nd}^\mathrm{full} = -2 B_3$.
\end{itemize}

\vspace{6pt}
In \texttt{allesfitter}, this phase curve model can be selected by setting \textit{phase\_curve\_style} to \textit{sine\_series} (see Table~\ref{tab:Settings}).
The above terms are parametrized with 
\begin{itemize}
    \item \textit{[companion]\_phase\_curve\_A1\_[inst]} for beaming,
    \item \textit{[companion]\_phase\_curve\_B1\_[inst]} for atmospheric,
    \item \textit{[companion]\_phase\_curve\_B2\_[inst]} for $1^\mathrm{st}$ ellipsoidal,
    \item \textit{[companion]\_phase\_curve\_B3\_[inst]} for $2^\mathrm{nd}$ ellipsoidal.
\end{itemize} (see Table~\ref{tab:Parameters}). We do not include the terms $A_2$ and $A_3$, which have no physical interpretation.

The atmospheric component can further be separated into thermal and reflected components, both of which can receive a phase shift, using the expanded set of parameters explained in Table~\ref{tab:Parameters}.

\subsubsection{Phase curves using transformed sines}
\label{sss:Phase curves using transformed sines}

A drawback with the option above is that the pure harmonic series of waveforms requires that some semi-amplitudes must be negative to admit a physical interpretation.
For example, a user might instead desire to fit for a `physical' full (peak-to-peak) amplitude of the atmospheric component.

By selecting the setting \textit{phase\_curve\_style} as \textit{sine\_physical} (see Table~\ref{tab:Settings}) the user can therefore model the phase curve with a linear combination of sinusoids while defining all quantities as physical quantities. The respective set of parameters is:
\begin{itemize}
    \item \textit{[companion]\_phase\_curve\_beaming\_[inst]}: positive semi-amplitude of the beaming effect, representing the term $A_1 \sin{\phi(t)}$, i.e. a modulation around the median flux level of the star.
    \item \textit{[companion]\_phase\_curve\_atmospheric\_[inst]}: positive full (peak-to-peak) amplitude of the atmospheric contribution, representing the term $-2 B_1 (1 - \cos{(\phi(t))}$, i.e. an additive component to the companion's nightside flux. 
    \item \textit{[companion]\_phase\_curve\_ellipsoidal\_[inst]}: positive full (peak-to-peak) amplitude of the leading-order term of the ellipsoidal modulation, representing the term $-2 B_2 (1 - \cos{(2\phi(t))})$, i.e. an additive component to the system's flux from spherical (non-distorted) bodies. 
    \item \textit{[companion]\_phase\_curve\_ellipsoidal\_2nd\_[inst]}: positive full (peak-to-peak) amplitude of the next-order term of the ellipsoidal modulation, representing the term $-2 B_3 (1 - \cos{(3\phi(t))})$, i.e. an additive component to the system's flux from spherical (non-distorted) bodies. 
\end{itemize}

\vspace{6pt}
As above, the atmospheric component can incorporate phase shifts and allows to distinguish between thermal and reflected contributions, using the expanded set of parameters explained in Table~\ref{tab:Parameters}.

\subsubsection{Phase curves with ellc's physical model}
\label{sss:Phase curves with ellc's physical model}

An alternative way to model these effects with \texttt{allesfitter} is utilizing \texttt{ellc}'s relevant physical forward-model directly by using the setting \textit{phase\_curve\_style} as \textit{ellc}.
The physical model is driven by parameters which affect the computation of the heated dayside of the companion, \textit{[companion]\_heat\_[inst]}, the gravity darkening coefficients, \textit{[companion]\_gdc\_[inst]}, and the Doppler boosting factor \textit{[companion]\_bfac\_[inst]}, as well as the desired stellar shape approximation (see Section~\ref{ss:Stellar/planetary grid and shape} and Tables~\ref{tab:Settings} and \ref{tab:Parameters}). 
As this approach requires a thorough understanding of the chosen settings and parameters, we only recommend it to users who are proficient with \texttt{ellc}. For a detailed description of all effects and caveats we thus refer the reader to \citet{Maxted2016}.

\subsection{Stellar/planetary grid and shape}
\label{ss:Stellar/planetary grid and shape}

The \texttt{ellc} implementation constructs all objects in the system as three dimensional bodies and computes the light curve and RV forward-models by integration over the visible surfaces. This allows a physically accurate representation of star spots and heat redistribution on the surface. The user can set the density of this interpolation grid using one of the five options from \textit{very\_sparse} to \textit{very\_fine} (see Table~\ref{tab:Settings}). The available grid options have a strong impact on the computational speed, but usually do not noticeably impact the results \citep[see][]{Maxted2016}. We thus recommend the user to run all test runs with \textit{very\_sparse}, and only run the publication-ready model with a finer spacing.

Additionally, the user can efficiently compute deviations of the stellar/planetary shape (see Table~\ref{tab:Settings}). The \textit{sphere} option is the default and appropriate for any model that do not incorportate interaction between the objects. The \textit{roche} shape calculates the object's shape using the Roche equation \citep{Wilson1979}. The \textit{roche\_v} shape is suited for synchronous rotation, where the volume of the star can be modeled via \citet{Kopal1978}. With \textit{poly1p5} or \textit{poly3p0} the object is modeled as a polytrope with index n=1.5 or n=3.0, respectively \citep{Chandrasekhar1933, James1964}. Finally, the \textit{love} option computes the objects' shape via \citep{Correia2014}.

\subsection{Derived parameters} 
\label{ss:Derived parameters} 
In addition to the fitted parameters, \texttt{allesfitter} also uses the samples drawn from the posterior distribution of parameters to derive an extensive list of additional quantities. The full list is shown in Table~\ref{tab:Derived parameters}, along with explanations on how these values are derived from the posterior samples.

\section{Examples and case studies}
\label{s:Examples and case studies}

\subsection{The two-planet system Pi~Mensae}

\begin{figure*}[!htbp]
    \centering
    \includegraphics[width=\textwidth]{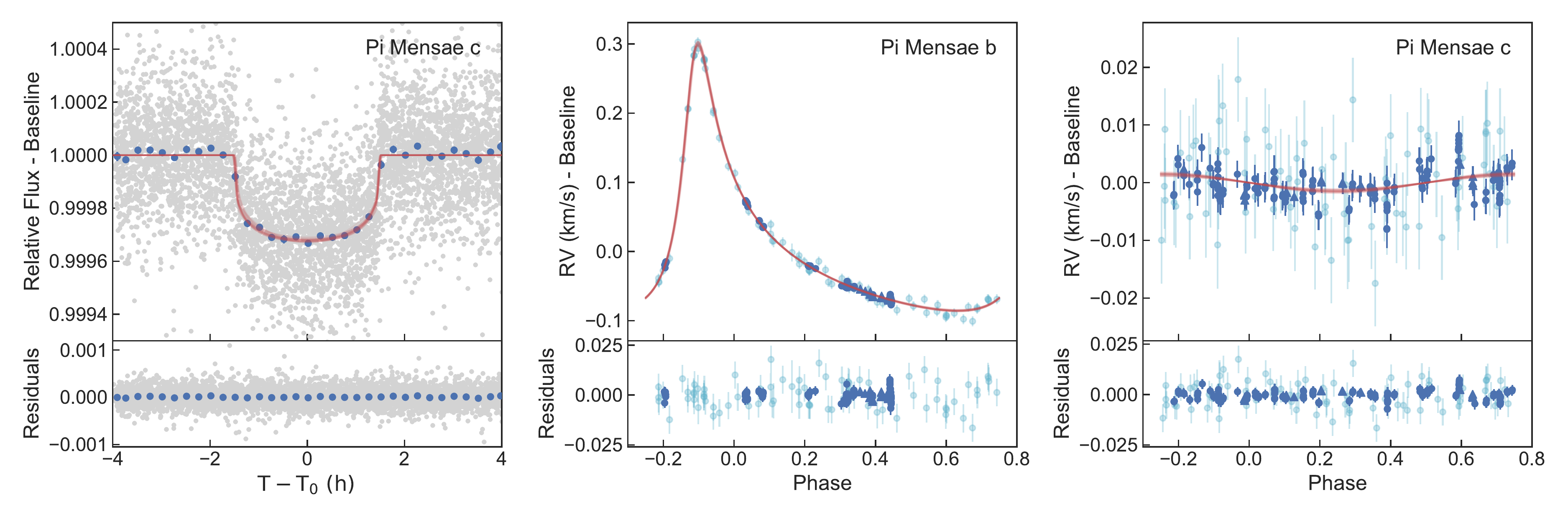}
    \caption{Global fit to the two planet system Pi~Mensae. The system hosts Pi~Mensae\,c, the first exoplanet discovered by \TESS{}. The shown data are from \TESS{}, AAT (light blue), and HARPS (blue). Red curves show 20 fair samples drawn from the posterior. The photometric red noise floor was estimated with a GP Mat{\'e}rn 3/2 kernel and removed from the data before phase-folding.}
    \label{fig:Pi_Mensae_fit}
\end{figure*}

\begin{figure}[!htbp]
    \centering
    \includegraphics[width=\columnwidth]{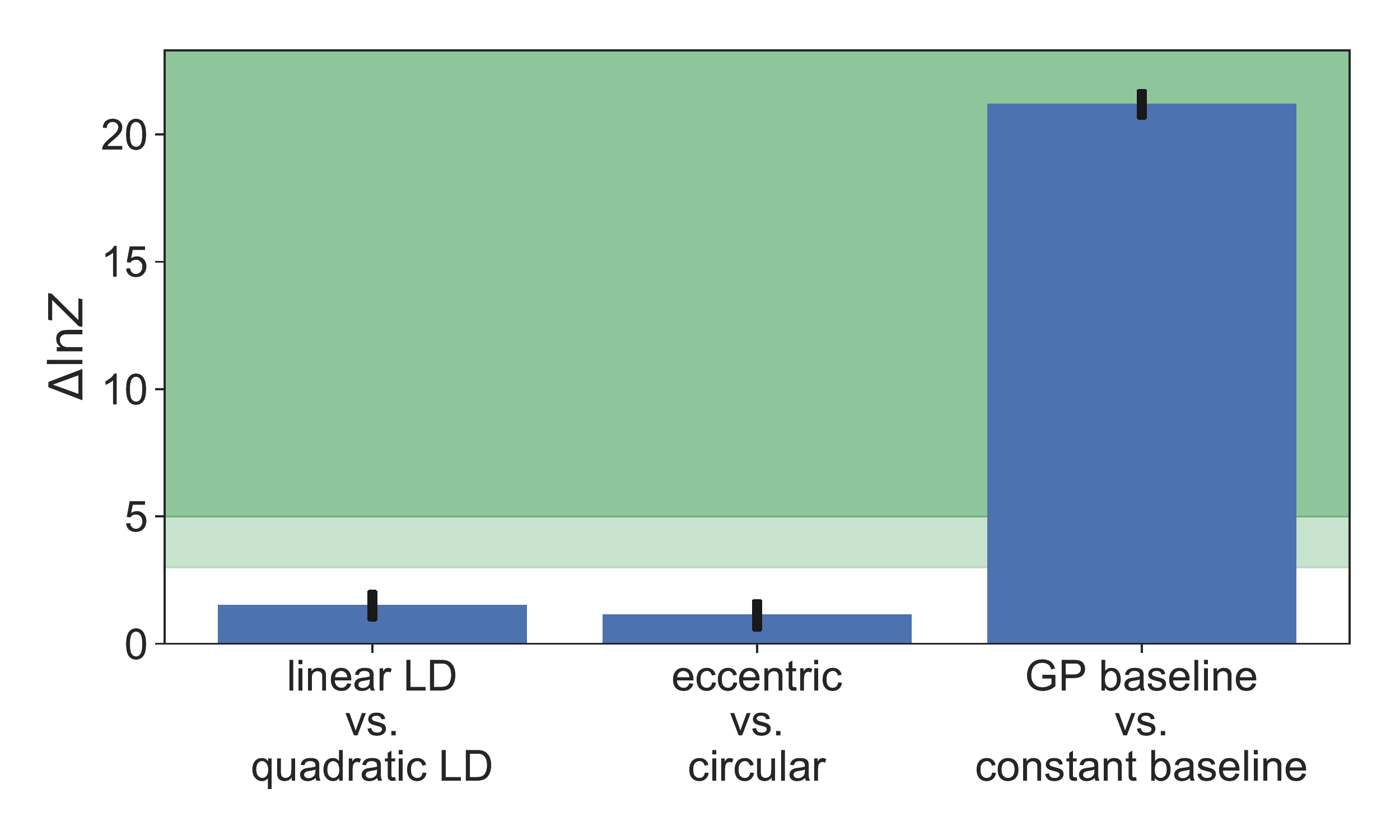}
    \caption{Bayes factors, $\Delta \log Z$, comparing different models for the Pi~Mensae system using \TESS{} Sector 1 and all RV data (comparable to \citet{Huang2018b}). 
    There is no strong evidence for/against a linear limb darkening model as opposed to a quadratic limb darkening model (left).
    Likewise, there is no strong evidence for an eccentric orbit over a circular orbit for planet c (middle).
    There is, however, strong evidence for using a GP Mat{\'e}rn 3/2 kernel over a constant baseline, indicative of short-term systematics that remained in the spline-detrended light curve (right).
    }
    \label{fig:Pi_Mensae_Bayes_factors}
\end{figure}

\begin{figure}[!htbp]
    \centering
    \includegraphics[width=\columnwidth]{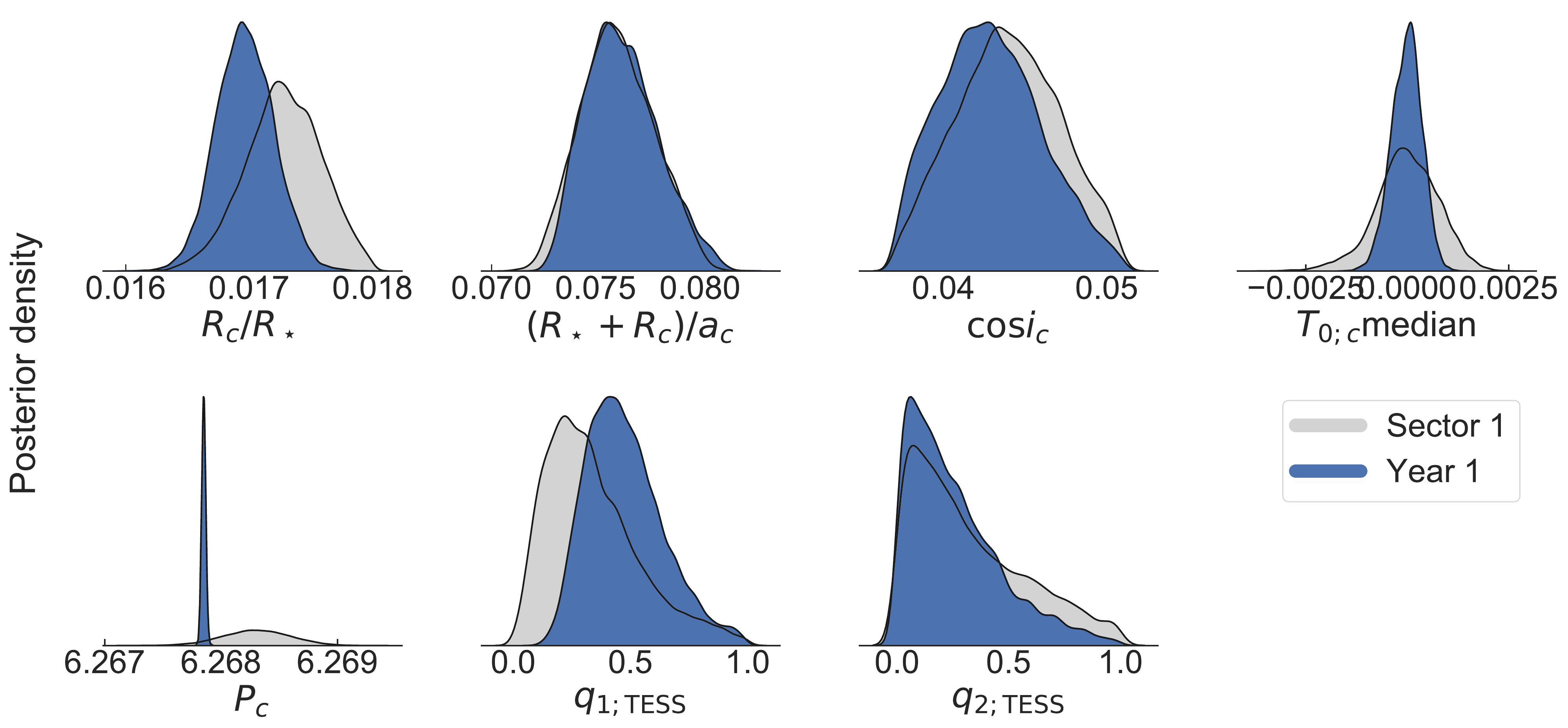}
    \caption{Updates to the posteriors of the major astrophysical parameters of Pi~Mensae\,c with new \TESS{} data. Grey shaded distributions show the results of our \texttt{allesfitter} analysis of Sector 1 data. Blue shaded distributions show the refinement we could achieve by adding all available data from \TESS{} Year 1 (Sectors 1, 4, 8, 11, 12, and 13). In particular, we find a significant improvement in the precision and accuracy of planet c's orbital period, which is a direct consequence of the longer observation baseline.}
    \label{fig:Pi_Mensae_posterior_histograms}
\end{figure}

In this section, we demonstrate how \texttt{allesfitter} can be used to infer the parameters of a multi-planet system from photometric and RV data from different telescopes. We also show how the Bayesian evidence can be used to compare different models for limb darkening laws, eccentric versus circular orbits, and systematic noise.

For this, we first re-analyze \TESS{}' first exoplanet discovery, Pi~Mensae\,c \citep[][hereafter H18]{Huang2018b}, using \TESS{} Sector 1 data only and compare our results with those from H18 (Section~\ref{sss:Re-analysis of TESS Sector 1 and RV data}). 
Afterwards, we include all new available data form \TESS{} Year 1 in an effort to update the literature values (Section~\ref{sss:New analysis of all TESS Year 1 and RV data})\footnote{all data, code, and results available at \url{https://github.com/MNGuenther/allesfitter/paper/Pi_Mensae}}.

\subsubsection{Re-analysis of \TESS{} Sector 1 and RV data}
\label{sss:Re-analysis of TESS Sector 1 and RV data}

The Pi~Mensae system hosts two known planets. The 10 Jupiter mass planet Pi~Mensae\,b was originally discovered using RV surveys \citep{Jones2002,Wittenmyer2012} on a highly eccentric 5.7 year orbit. In 2018, shortly after its launch, \TESS{} unveiled photometric transits of an inner companion, Pi~Mensae\,c, only twice the size of Earth and on a 6.27 day orbit. We have already used \texttt{allesfitter} for an independent analysis in the \TESS{} discovery paper, and here we showcase our results in more depth, emphasizing additional aspects.

We perform five different re-analyzes of the Sector 1 data from H18 in global fits of all available photometric and RV data, i.e., their \TESS{} Sector 1 spline-detrended light curve along with all archival RV data from HARPS and AAT. We define the following as our `standard' settings to reproduce the original study by H16: we use constant baselines for the \TESS{} and RV data, assume a circular orbit for c, and apply a quadratic limb darkening law. The only difference to H16 is that we uniformly sample in the transformed parameter space from \citet{Kipping2013}, while the original study set the quadratic limb darkening to tabulated values from \citep{Claret2017}.

We then compare the results and Bayesian evidences (i.e., using Nested Sampling) for different variations of the above standard settings:
\begin{enumerate}
\item MCMC with the standard settings,
\item Nested Sampling with the standard settings,
\item Nested Sampling with free eccentricity for planet c,
\item Nested Sampling with a linear limb darkening law,
\item Nested Sampling with a GP Mat{\'e}rn 3/2 baseline for the \TESS{} data.
\end{enumerate}

In all approaches, we uniformly sample from the posterior of the radius ratios $R_\mathrm{b}/R_\star$ and $R_\mathrm{c}/R_\star$, 
sums of radii over semi-major axis $(R_\mathrm{b} + R_\star)/a$ and $(R_\mathrm{c} + R_\star)/a$,
cosines of the inclination $\cos{i_\mathrm{b}}$ and $\cos{i_\mathrm{c}}$,
eccentricity and argument of periastron as $\sqrt{e_\mathrm{b}} \cos{\omega_\mathrm{b}}$, $\sqrt{e_\mathrm{b}} \sin{\omega_\mathrm{b}}$,
quadratic limb darkening in the \citet{Kipping2013} transformation $q_1$ and $q_2$,
a constant baseline offset $\Delta F$,
and the white noise error scaling $\ln \sigma_F$.
For variation 2), we also uniformly sample from the posterior of $\sqrt{e_\mathrm{c}} \cos{\omega_\mathrm{c}}$ and $\sqrt{e_\mathrm{c}} \sin{\omega_\mathrm{c}}$.
For variation 3), we sample from the posterior of linear limb darkening instead of quadratic.
For variation 4), we first fit the GP Mat{\'e}rn 3/2 model to the out-of-transit data, and then apply normal priors on it for the fit to the in-transit data. \citep[see e.g.][]{Guenther2019b}.

We find a good fit to the data (Fig.~\ref{fig:Pi_Mensae_fit}), and all results from our different model variations agree well with one another and with those published by H18; MCMC and Nested Sampling give consistent results for the standard settings. 
Furthermore, comparing the Bayesian evidences of all Nested Sampling model fits, we find that the model with a GP Mat{\'e}rn 3/2 baseline is strongly favored (Fig.~\ref{fig:Pi_Mensae_Bayes_factors}). This is likely because the \TESS{} Sector 1 data from H18 were affected by remnant systematics on time scales shorter than those removed by the original spline detrending. These short term systematics are possibly caused by the pointing jitter of the satellite, which is now well characterized and understood. 
Moreover, we find that the circular orbit assumption for planet c and the choice of a quadratic limb darkening model are justified by the data.

\subsubsection{New analysis of all \TESS{} Year 1 and RV data}
\label{sss:New analysis of all TESS Year 1 and RV data}

Finally, we go beyond a mere comparison with the discovery paper and update the literature values for Pi~Mensae by analyzing all available \TESS{} data from the first year of operations, i.e. observations from Sectors 1, 4, 8, 11, 12, and 13, along with all RV data used in H18. Due to the bright host star, we use custom-aperture light curves which are detrended against the quaternions and the first 7 components of the co-trending basis vectors (custom light curves courtesy of Chelsea X. Huang).
Our \texttt{allesfitter} approach is equivalent to variation 5 in Section~\ref{sss:Re-analysis of TESS Sector 1 and RV data} (i.e. circular orbit of planet c; quadratic limb darkening; GP baseline; Nested Sampling). 
The resulting fit is shown in Fig.~\ref{fig:Pi_Mensae_fit}, and all results are summarized in Table~\ref{tab:Pi_Mensae_results} and Fig.~\ref{fig:Pi_Mensae_ns_corner}.

\input{tab_Pi_Mensae_results}

We again find a good agreement with the discovery paper, along with a significant improvement in the median and precision of planet c's orbital period, as expected from the extended observing baseline (see Fig.~\ref{fig:Pi_Mensae_posterior_histograms}).
We also find that the updated detrending of the full \TESS{} Year 1 light curve, now incorporating all state-of-the-art understanding of systematics, removed the remnant short-term noise which was picked up by the GP in Section~\ref{sss:Re-analysis of TESS Sector 1 and RV data}. Hence, for the Year 1 analysis, the GP baseline turns out flat and is comparable to a constant offset. This also marginally updates our posteriors of the radius ratio and limb darkening.

\subsection{TTVs in the TOI-216 system}

\texttt{Allesfitter} allows to fit a global light curve model with individual transit/eclipse mid-time offsets for each transit event, even if those occur for multiple companions and were observed by different telescopes. 
We highlight these abilities on the example of the two-planet system TOI-216 (TIC 55652896), the first discovery by \TESS{} that shows clear transit timing variations \citep[TTVs;][]{Dawson2019, Kipping2019}.
From only the first few months of \TESS{} data, the system has been characterized to contain a pair of warm, large exoplanets. 
These planets orbit at mean-periods near 17.1 and 34.5 days, close to a 2:1 mean-motion resonance. 
\citep{Dawson2019}, in particular, analyze the \TESS{} Sectors 1-6 TTVs and find two families of solutions for the masses of planet b and c, respectively: either like a sub-Saturn and Neptune, or like a Jupiter and sub-Saturn.

Here, we analyze TOI-216 with \texttt{allesfitter} while freely fitting for the transit mid-times, with the goal of deriving all planetary and orbital parameters including a TTV O-C diagram (i.e. observed minus calculated)\footnote{all data, code, and results available at \url{https://github.com/MNGuenther/allesfitter/paper/TOI-216}}. 
We include a total of 12 Sectors of \TESS{} data, which have been collected for this target by now (Sectors 1-9 and 11-13), doubling the original baselines of the discovery papers.

We uniformly sample from the posterior of the radius ratios $R_\mathrm{b}/R_\star$ and $R_\mathrm{c}/R_\star$, 
sums of radii over semi-major axis $(R_\mathrm{b} + R_\star)/a$ and $(R_\mathrm{c} + R_\star)/a$,
cosines of the inclination $\cos{i_\mathrm{b}}$ and $\cos{i_\mathrm{c}}$,
quadratic limb darkening in the \citet{Kipping2013} transformation $q_1$ and $q_2$,
a GP Mat{\'e}rn 3/2 baseline with parameters $\ln \rho_\mathrm{GP}$ and $\ln \sigma_\mathrm{GP}$,
and the white noise flux error scaling $\ln \sigma_F$.
We first fit the GP Mat{\'e}rn 3/2 model to the out-of-transit data, and then apply normal priors on it for the fit to the in-transit data. \citep[see e.g.][]{Guenther2019b}.

Notably, the grazing transit of planet b leads to a degeneracy between the radius ratio and orbital inclination, which can lead to a `runaway' solution if using wide uniform priors and no external constrains.
A possible way to overcome this is by implying an external planet density prior \citep[e.g.][]{Bayliss2018NGTS-1b:M-dwarf}. 
For this example, however, we chose to follow the approach by \citet{Dawson2019} and constrain the radius ratio to a uniform prior between 0 and 0.17, since the `runaway' solution starts around $>$0.2.

\begin{figure}[!htbp]
    \centering
    \includegraphics[width=\columnwidth]{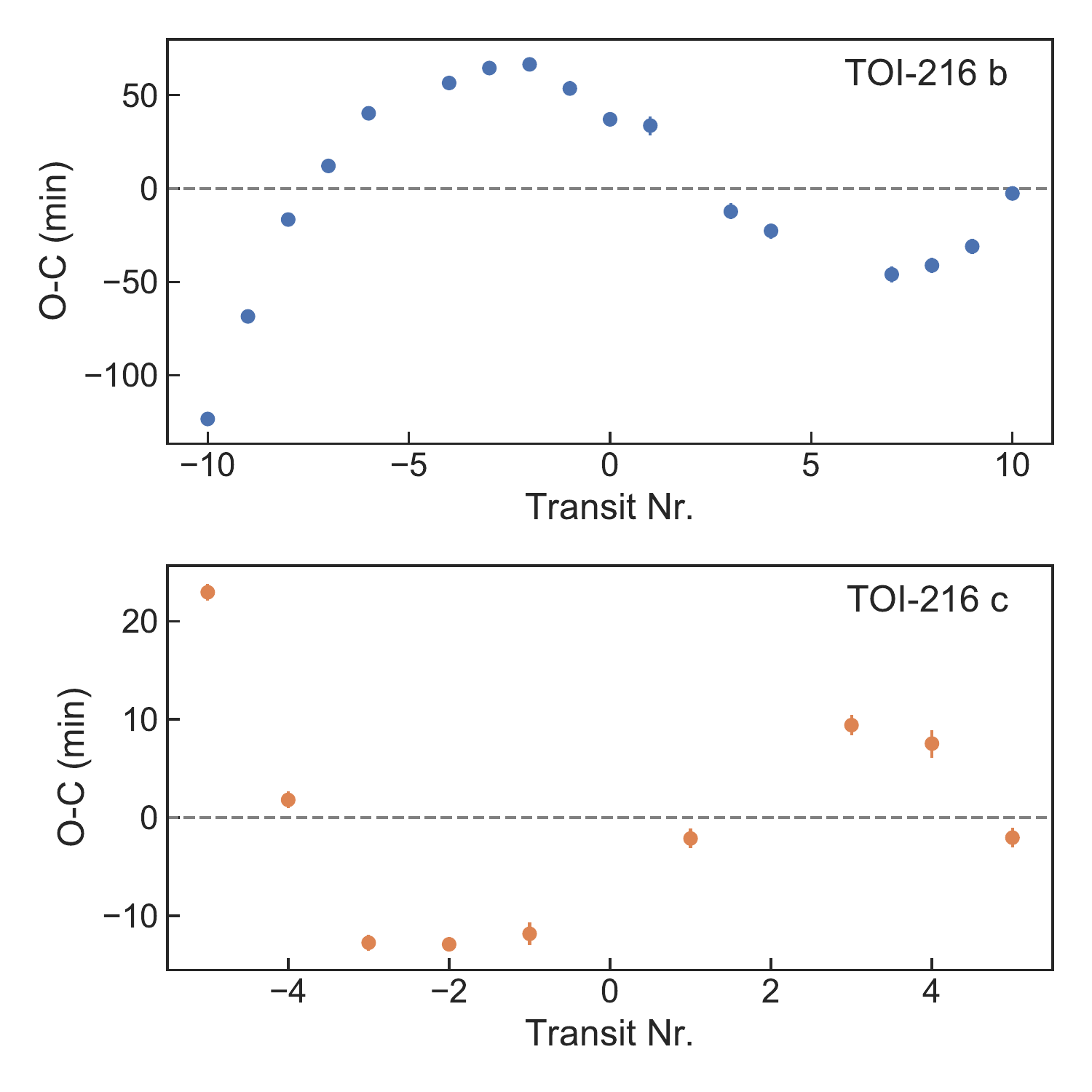}
    \caption{Updated TTV O-C diagram for TOI-216 from all \TESS{} Year 1 data (Sectors 1-9 and 11-13). The system hosts two warm, large exoplanets near a mean-motion resonance of 2:1. The O-C diagrams were created by removing a linear trend from the posterior transit mid-times. The curves show that nearly one complete TTV super-period has been sampled.}
    \label{fig:TOI-216_o_minus_c}
\end{figure}

\begin{figure*}[]
    \centering
    \includegraphics[width=0.6\textwidth]{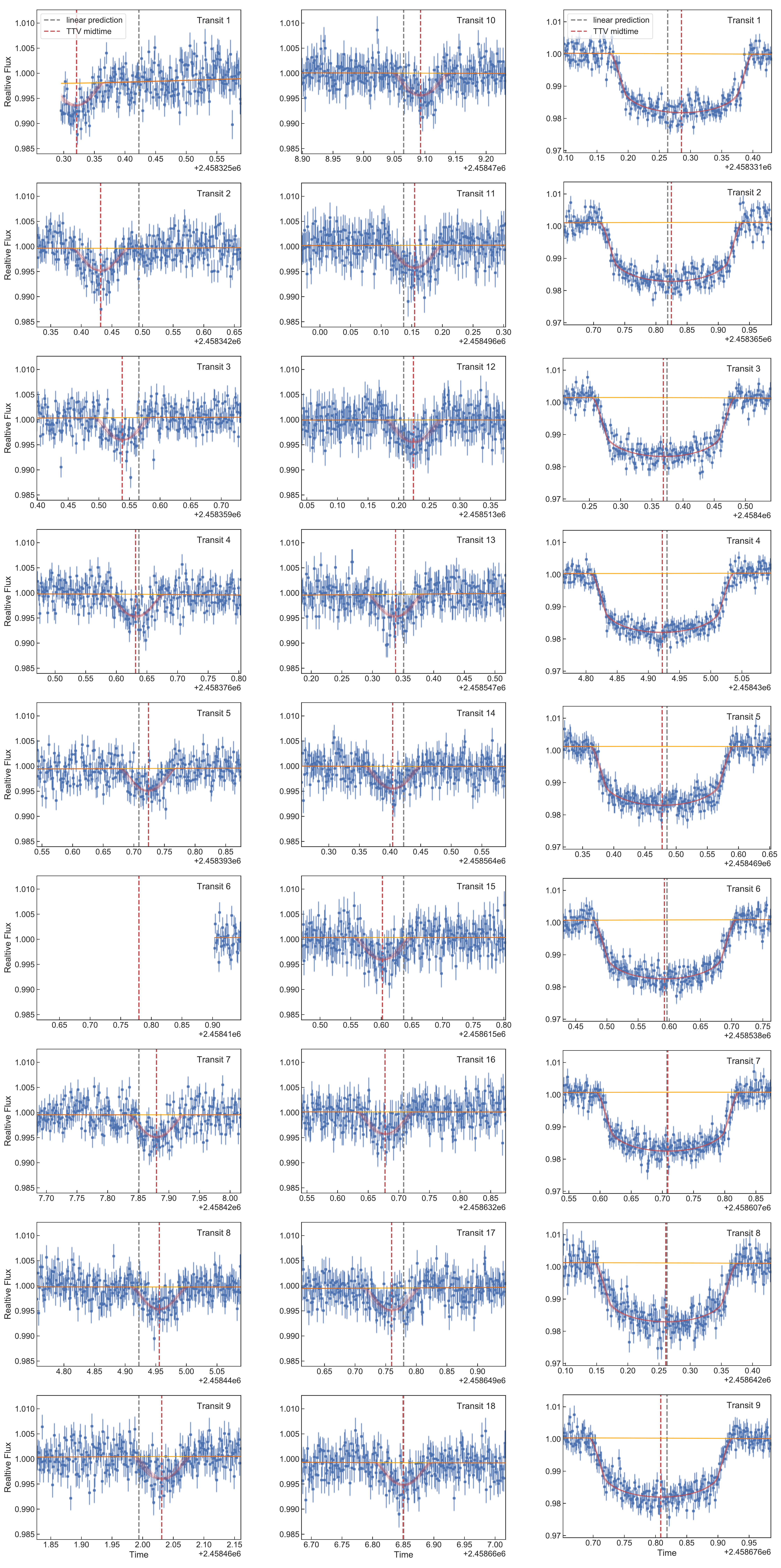}
    \caption{Global fit with free TTVs to the TOI-216 system from all \TESS{} Year 1 data (Sectors 1-9 and 11-13). The system hosts two warm, large exoplanets near a mean-motion resonance of 2:1. All shown data are \TESS{} 2-minute cadence observations. The left and middle column show transit windows for TOI-216\,b, the right column those for TOI-216\,c. Red curves show 20 fair samples drawn from the posterior of the global model including free TTVs for each transit. Orange curves show 20 fair samples drawn from the posterior of the GP Mat{\'e}rn 3/2 baseline model.}
    \label{fig:TOI-216_ns_fit_per_transit}
\end{figure*}

We find a good fit to the data, and our results agree well with those form \citet{Dawson2019} and \citet{Kipping2019}. 
The per-transit light curves and posterior models are shown in Fig.~\ref{fig:TOI-216_ns_fit_per_transit}.
By including all available \TESS{} Year 1 data and hence doubling the baseline from \citet{Dawson2019} and \citet{Kipping2019}, we can also update the TTV O-C diagrams, as shown in Fig.~\ref{fig:TOI-216_o_minus_c}.
All posteriors are summarized in Table~\ref{tab:TOI-216_results} for updated physical and orbital parameters, Table~\ref{tab:TOI-216_o_minus_c} for updated transit mid-times and TTV O-C values, and Fig.~\ref{fig:TOI-216_ns_corner_compressed} for posterior corner plots. 

\input{tab_TOI-216_results}

\input{tab_TOI-216_o_minus_c}

\newpage
\subsection{The phase curve of WASP-18b}

\texttt{Allesfitter} can also model phase curves of exoplanets and binary stars, which we demonstrate here on the example of the hot Jupiter WASP-18\,b (TIC 100100827) \citep{Hellier2009, Southworth2009}. The system harbors a 10 Jupiter mass companion on a short orbital period of 0.94 days. This extreme combination leads to interactions between the star and planet that cause a phase curve signature at visible wavelengths. In turn, studying this phase curve gives insight into the atmosphere of this hot Jupiter. As for Pi~Mensae (see above), \texttt{allesfitter} was already used to perform an independent analysis for the original \TESS{} study by \citet{Shporer2019} (hereafter S19), and we here showcase how such an analysis can be performed.

\begin{figure*}[!htbp]
    \centering
    \includegraphics[width=\textwidth]{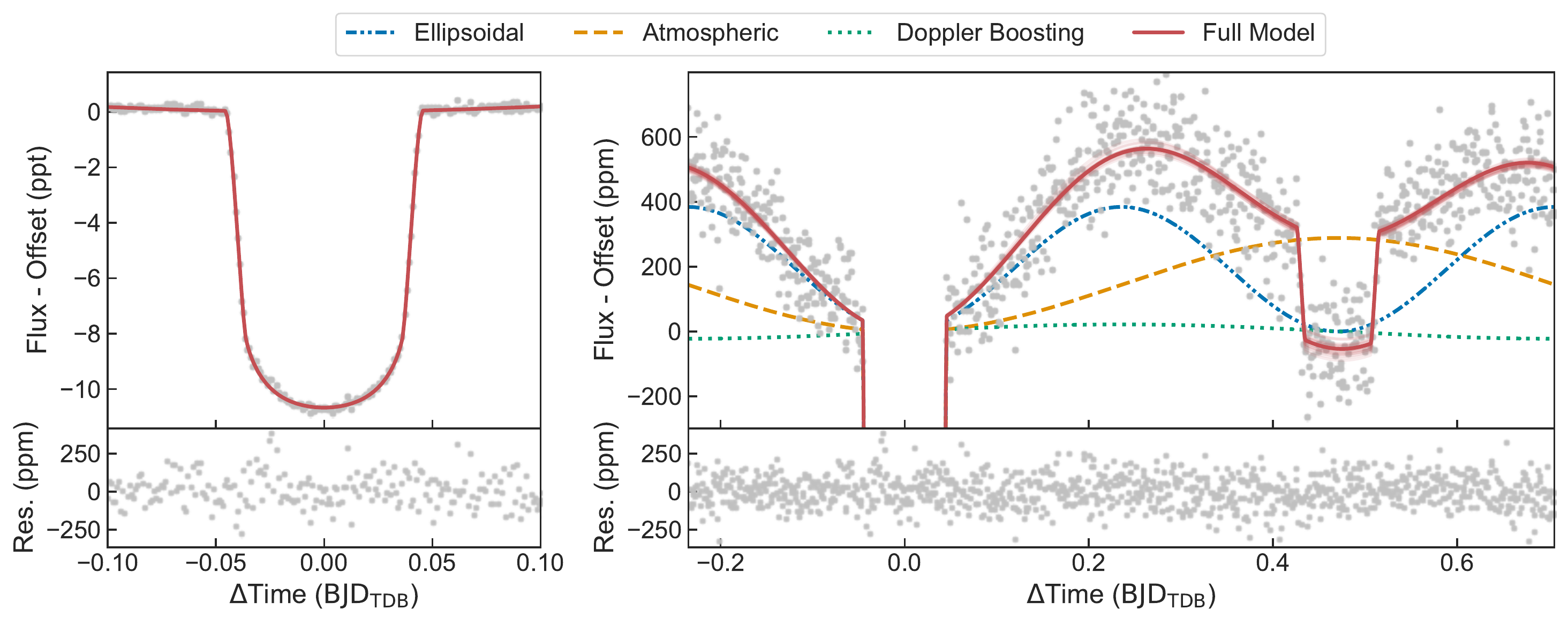}
    \caption{Global fit to the \TESS{} phase curve of the hot Jupiter WASP-18\,b (grey points). 
    Lines show the models generated from the posterior median, for the ellipsoidal modulation (blue dot-dashed), atmospheric component (orange dashed), Doppler boosting / beaming (green dotted) and the full model including transit and occultation (red line) using \texttt{allesfitter}'s sine\_physical phase curve model.
    Semi-transparent red lines show 20 full models randomly drawn from the posteriors.}
    \label{fig:WASP-18_fit}
\end{figure*}

For this example, we speed up our analysis by phase-folding the \TESS{} light curve on an epoch of 2458361.048072 $\mathrm{BJD_{TDB}}$ and
period of 0.9414576\,days, which are posterior medians from our preliminary analysis. We then bin the phase curve over a grid of 1000 points in phase, which corresponds to a bin width of 1.4\,min. 
We perform two fits, one with the `sine\_series' phase curve model (as in S19) and the other with the `sine\_physical' phase curve model.
We uniformly sample from the posterior of the radius ratio $R_\mathrm{b}/R_\star$, 
sum of radii over semi-major axis $(R_\mathrm{b} + R_\star)/a$,
cosine of the inclination $\cos{i}$,
surface brightness ratio $J$ of the planet's dayside and star,
the Doppler boosting (beaming) effect ($A_1$ in sine\_series, $A_\mathrm{beaming}$ in sine\_physical),
the atmospheric modulation from thermal emission and reflected light ($B_1$ in sine\_series, $A_\mathrm{atmospheric}$ in sine\_physical),
the ellipsoidal modulation ($B_2$ in sine\_series, $A_\mathrm{ellipsoidal}$ in sine\_physical),
a constant baseline offset $\Delta F$,
and the white noise error scaling $\ln \sigma_F$.
As we here only fit photometric \TESS{} data, we also apply prior information from RV observations.  For simplicity in this example, we fix the eccentricity to $e=0.0091$ and
argument of periastron to $\omega=269\,^\circ$ \citep{Knutson2014, Stassun2017}.
We run an MCMC analysis starting from the values found by previous studies, with 500 walkers, a thinning of 50 steps, 1000 burn-in steps and 5000 total steps. We consider the fits to be converged as all chains are $>$42{}$\times$ their autocorrelation lengths.

We find a good fit to the \TESS{} light curve of WASP-18 (Fig.~\ref{fig:WASP-18_fit}) and a good agreement with the results from S19 (Fig.~\ref{fig:WASP-18_mcmc_corner}) with both phase curve settings. In particular, we can individually interpret the components of the phase curve forward-model. 
The ellipsoidal modulation in our sine\_series model has a semi-amplitude of $-192.2\pm5.9$\,ppm ($-190.5_{-5.9}^{+5.8}$\,ppm in S19).
We also find evidence for Doppler boosting, with a semi-amplitude of $22.1\pm4.5$\,ppm ($21.0\pm4.5$\,ppm in S19).
There is a slight difference in our semi-amplitude of the atmospheric phase modulation $-144.3\pm5.6$\,ppm and radius ratio of $0.09757\pm0.00014$ compared to S19 ($-174.4_{-6.2}^{+6.4}$\,ppm and $0.09716_{-0.00014}^{+0.00014}$, respectively). 
This is likely caused by our simplified example (phase-folded and binned data, fixed parameters, constant offset baseline) and the fact that S19 also fit a polynomial background model and additional higher-order sinusoidal harmonics.
Fitting for the surface brightness ratio of the planet's dayside and star, we find $J=0.0056\pm0.0016$, an occultation depth of $342\pm15$\,ppm ($341_{-18}^{+17}$\,ppm in S19).

\newpage
\subsection{The spotted binary star system KOI-1003}

\begin{figure*}[!htbp]
    \centering
    \includegraphics[width=\textwidth]{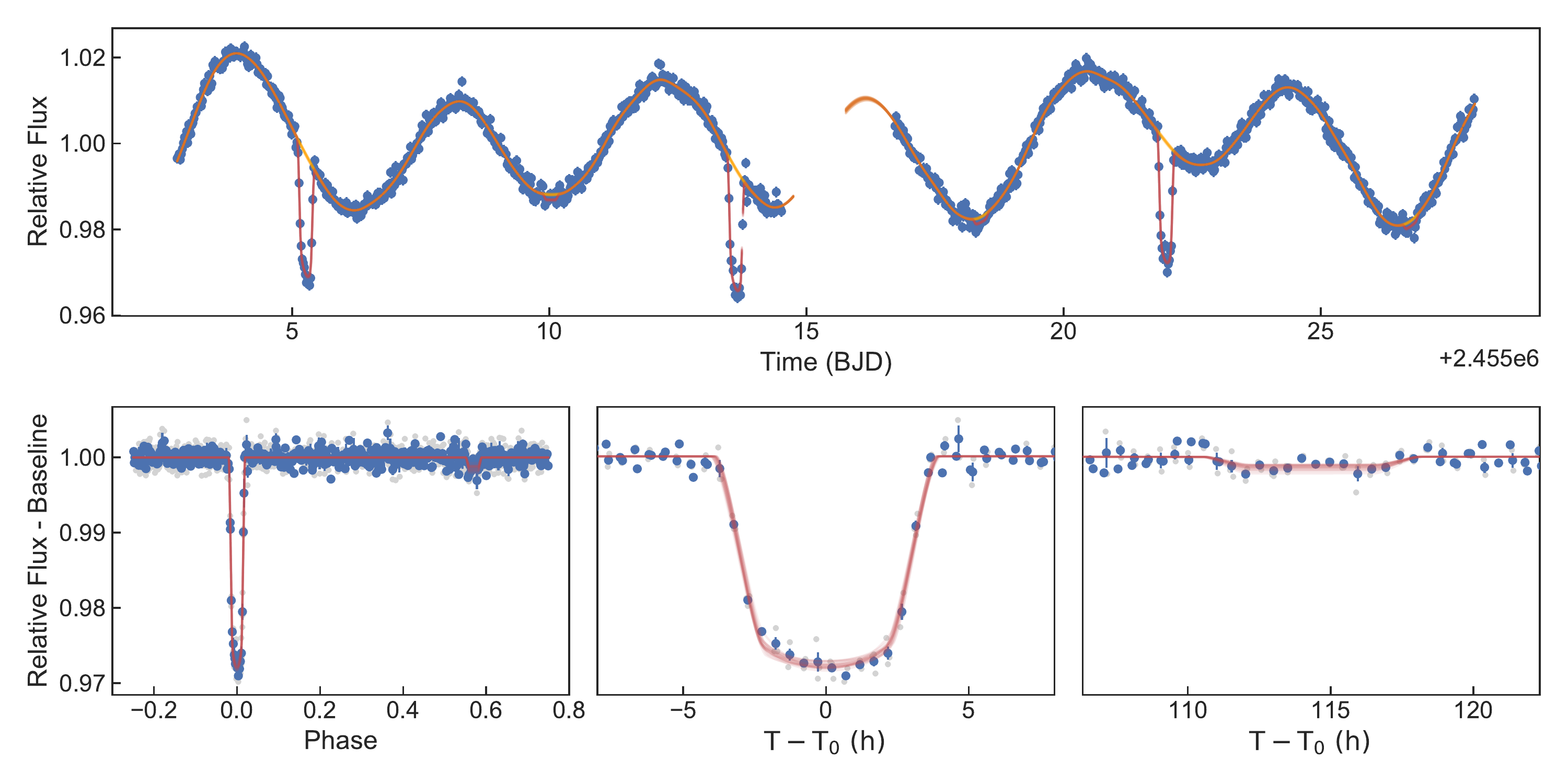}
    \caption{KOI-1003 model fit with stellar variability, overplotted with a 28\,day snapshot of the \Kepler{} long-cadence light curve of this spotted binary system (blue points).
    Red and orange lines show forward-model light curves generated using 20 fair draws from the posterior of the eclipse and GP models.}
    \label{fig:KOI-1003_fit}
\end{figure*}

In this example, we show how \texttt{allesfitter} can be used to infer parameters for binary star models in the presence of stellar variability and for long cadence data, using the example of KOI-1003 (TIC 122374527)\footnote{all  data,  code,  and  results  available  at  \url{https://github.com/MNGuenther/allesfitter/paper/KOI-1003}}.
KOI-1003 is an active, spotted binary star system \citep{Roettenbacher2016} (herafter R16) and is classified as a RS Canum Venaticorum (RS CVn) binary. Such systems are close binaries, where the primary is an evolved giant or sub giant that partially fills its Roche-potential and the secondary is a fainter main-sequence.
The star was observed in Kepler Quarters 2-17 nearly continuously with 29.4\,min cadence. R16 found that the binary's orbital and stellar rotation periods are nearly synchronized at 8.36 and 8.23\,days, respectively.
To create a fast-running example for the user, we here only utilize data from the first 28\,days of Quarter 2, covering three primary eclipses and three stellar rotation periods. 

We use this example to illustrate an approach of tackling similar system in multiple steps. 
In KOI-1003, the stellar variability is the dominant component of the observed light curve. To model it, we use \texttt{allesfitter} to mask out the eclipse regions and to fit a simple harmonic oscillating (SHO) GP along with the white noise scaling (see Section~\ref{ss:Baselines (red noise)}).
As initial guesses for the white noise scaling and SHO frequency we use \Kepler{}'s median flux error and $2\pi/(8.23/2)$\,days, respectively. We use half the rotation period, as two large opposite spots are apparent in the light curve. The initial guesses for the SHO amplitude and damping factor are set to small values, enforcing a smooth GP as the starting point for the MCMC. Our \texttt{allesfitter} run uses 500 walkers and performs one preliminary run with only 1000 steps in order to obtain relatively high-likelihood initial guesses for the nominal run. It then runs 1000 steps of burn-in and 5000 total steps, all thinned by a factor of 10, leading to 20,000 samples after convergence ($>$47$\times$ auto-correlation length).

Second, we utilize the trained GP to remove the stellar variability from the light curve. In the detrended light curve, we investigate if the shallow secondary eclipse can be detected despite the short range of data (expected depth of 1.8\,ppt from R16). To this end, we use \texttt{allesfitter}'s interface to call the \texttt{transit least squares} algorithm \citep{Hippke2019a}. We detect the primary eclipse with a period of 8.36\,days and depth of $\sim$28\,ppt at an SNR=38.2, and a second signal with a period of 8.7\,days and depth of $\sim$2\,ppt at SNR=5.3. We consider this to be likely related to the secondary eclipse, which is only a weak siggnal given the short range of data.

Third, we use the information gained above and perform a full model fit of the system with MCMC.
We uniformly sample from the posterior of the radius ratio $R_\mathrm{B}/R_\mathrm{A}$, 
sum of radii over semi-major axis $(R_\mathrm{B} + R_\mathrm{A})/a$,
cosine of the inclination $\cos{i}$,
quadratic limb darkening in the \citet{Kipping2013} transformation $q_1$ and $q_2$,
surface brightness ratio $J$,
the GP hyper-parameters,
and the white noise error scaling $\ln \sigma_F$.
We fix the eccentricity and argument of periastron as $\sqrt{e} \cos{\omega}$ and $\sqrt{e} \sin{\omega}$ to the values provided by R16, as our short data range does not reliably constrain the secondary eclipse.
We set the initial guesses for the physical values close to those by R16, and those for the GP and white noise scaling to the posterior medians obtained in the first step.
Since we analyze long-cadence data (29.4\,min), we also use a ten times finer evaluation grid to interpolate each point.
The MCMC is run with 500 walkers, 1000 steps of pre-run, 2000 steps of burn-in and 10000 total steps, all thinned by a factor of 100, leading to 40,000 samples after convergence ($>$33$\times$ auto-correlation length).

We find a good fit to the data which, despite the short data range, agrees well with the results from R16 (Figs.~\ref{fig:KOI-1003_fit} and \ref{fig:KOI-1003_mcmc_corner}). In particular, we find a period of $8.35992\pm0.00094$ ($8.360613\pm0.000003$ in R16), 
inclination of $85.75\pm0.31\,^\circ$ ($86.0\pm0.5$) and
ratio of semi-major axis to primary radius of $8.23\pm0.19$ ($8.2\pm0.5$ in R16).
We do find a significantly lower radius ratio of $0.1634_{-0.0021}^{+0.0017}$ ($0.177\pm0.003$ in R16) in this particular region of data, which is likely caused by spot crossings, i.e., the alignment of the planet with the stellar spots during the transit. This agrees with the fact that R16 found individual transit depths to vary between 2.73\% and 4.59\% due to spot crossings. Going forward, our modeling of these spot crossing could be refined by using a physical spot model (demonstrated in Section~\ref{ss:Star spots and flares on GJ 1243}) or including an additional short-term GP, e.g. using a Mat{\'e}rn 3/2 kernel.

Most importantly, by directly fitting for the surface brightness ratio of the eclipsing binary using a physical forward model, we find $J=0.053\pm0.012$, which could be used to constrain the spectral type of the secondary. The derived secondary eclipse depth of $1.40\pm0.31$\,ppt agrees well with R16 ($1.76\pm0.12$\,ppt in R16) and confirms the detection of the secondary eclipse.

\subsection{Star spots and flares on GJ 1243}
\label{ss:Star spots and flares on GJ 1243}

In addition to modeling eclipses of stars and exoplanets, \texttt{allesfitter} also models star spots and stellar flares. While star spots can cause semi-sinusoidal variations in the light curve as fainter regions rotate in and out of the visible disk, stellar flares cause an abrupt rise and subsequent exponential decay in the stellar brightness. A joint modeling of these effects can be relevant when stars exhibit both processes simultaneously, as often is the case for active M dwarfs.

Here, we demonstrate \texttt{allesfitter}'s abilities on the example of GJ~1243\footnote{all data, code, and results available at \url{https://github.com/MNGuenther/allesfitter/paper/GJ_1243}}. This M4 dwarf star is one of the most frequently flaring stars known, and was extensively studied with Kepler data \citep{Davenport2014,Hawley2014,Davenport2015,Silverberg2016}. These studies found evidence for a 0.59\,day rotation period, differential rotation and star spot evolution in four years of \Kepler{} data, along with a high flare frequency. \TESS{} recently re-observed the system during its Sector 14 and 15.

We analyze a 1.8\,day (three rotation periods) snapshot of \TESS{} observations, and simultaneously fit for star spots and stellar flares in this part of the light curve. 
We again use different approaches and compute the Bayesian evidence to compare the models:
\begin{itemize}
    \item two star spots and three flares,
    \item one star spot and three flares,
    \item two star spots and two flares.
\end{itemize}

Using Nested Sampling, we uniformly sample from the posterior of 
the rotation period,
star spots longitudes, latitudes, relative brightness, and sizes,
flares' peak times, amplitudes and FWHMs,
white noise scaling, 
and a constant baseline.
\citet{Silverberg2016} reported a spectroscopic $v \sin{i} \approx 25$\,km/s and stellar radius of $R_\star \approx 0.36\,\Rsun$. Using also the photometric rotation period $P_\mathrm{rot} \approx 0.59$\,days, we can compute
\begin{equation}
    i = \sin^{-1} \left(  \frac{v \sin{i} P_\mathrm{rot}}{2 \pi R_\star}  \right) \approx 54\,^\circ,
\end{equation}
where we freeze the inclination in our fit\footnote{note that the 31\,$^\circ$ reported in \citet{Silverberg2016} and \citet{Davenport2015} are apparently erroneous, and should have been $54\,^\circ$}.

\begin{figure*}[!htbp]
    \centering
    \includegraphics[width=\textwidth]{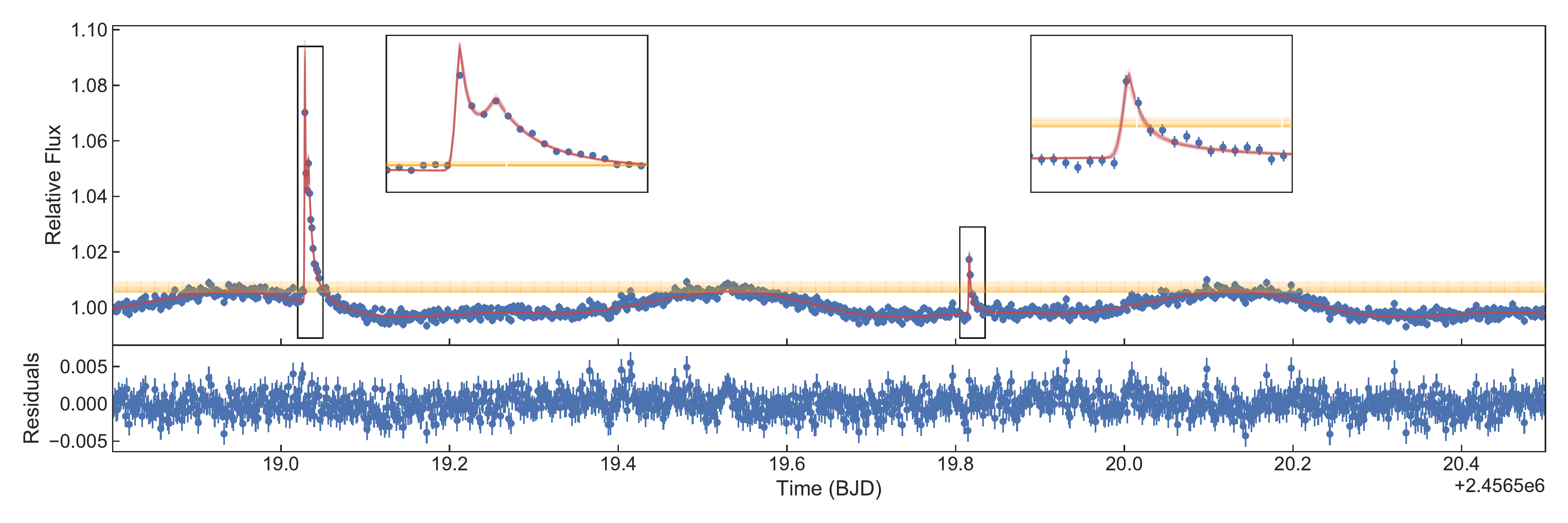}
    \caption{GJ 1243 model fit with two star spots and three stellar flares. Shown is a 1.8\,day snapshot of the \TESS{} Sector 14 data with 2 minute cadence (blue points), along with 20 randomly drawn \texttt{allesfitter} posterior samples for the baseline (orange lines) and full physical model (red lines). Inset plots show an enlarged view onto the flares.}
    \label{fig:GJ_1243_fit}
    
    \centering
    \includegraphics[width=\columnwidth]{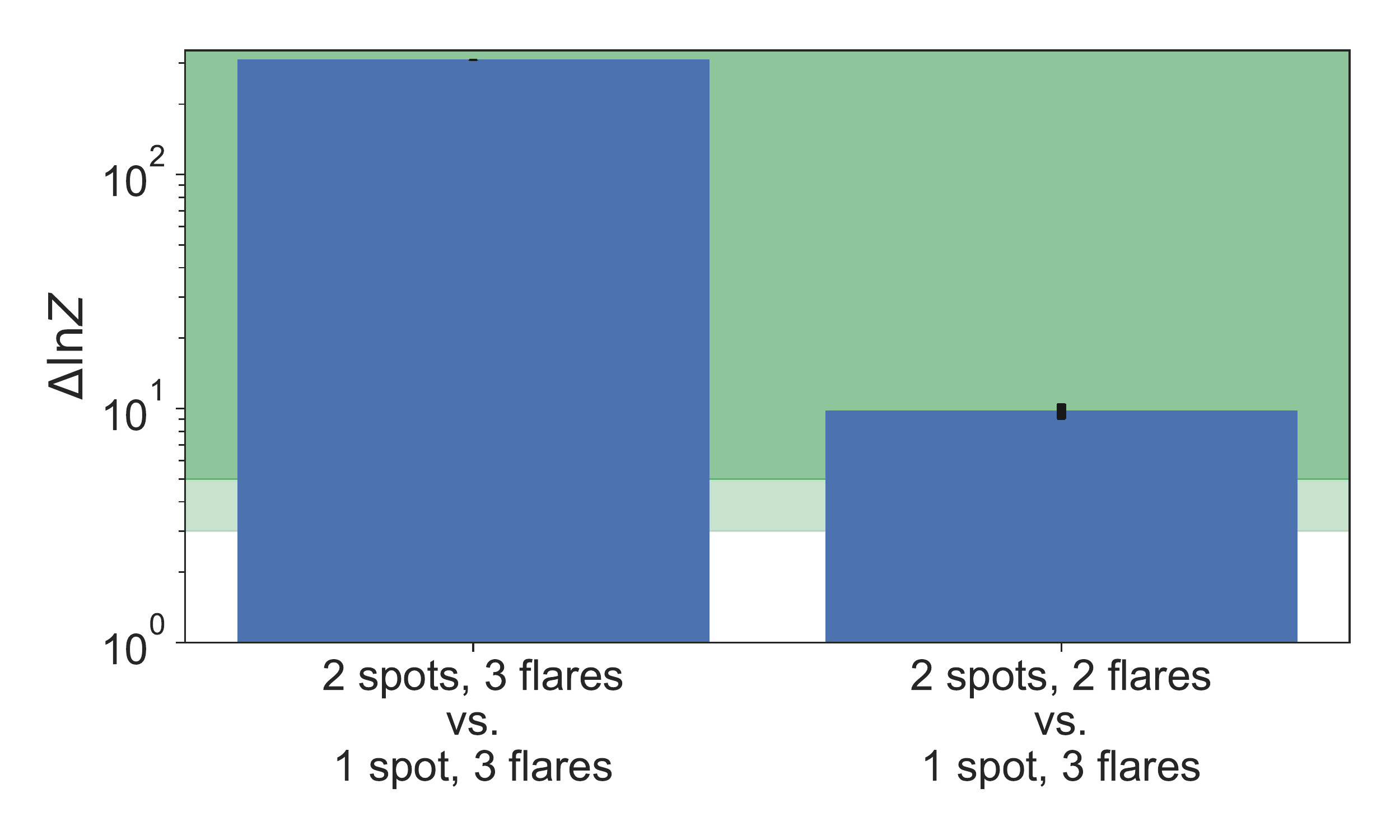}
    \caption{Bayes factors comparing the different models for GJ 1243. The model with 2 spots and 3 flares is clearly favored over the other models. Note that the 2 spot model is so strongly favored that a logarithmic y-axis scaling is needed for visualisation (i.e., on top of the already logarithmic $\Delta \log Z$).}
    \label{fig:GJ_1243_Bayes_factors}
    
    \centering
    \includegraphics[width=\columnwidth]{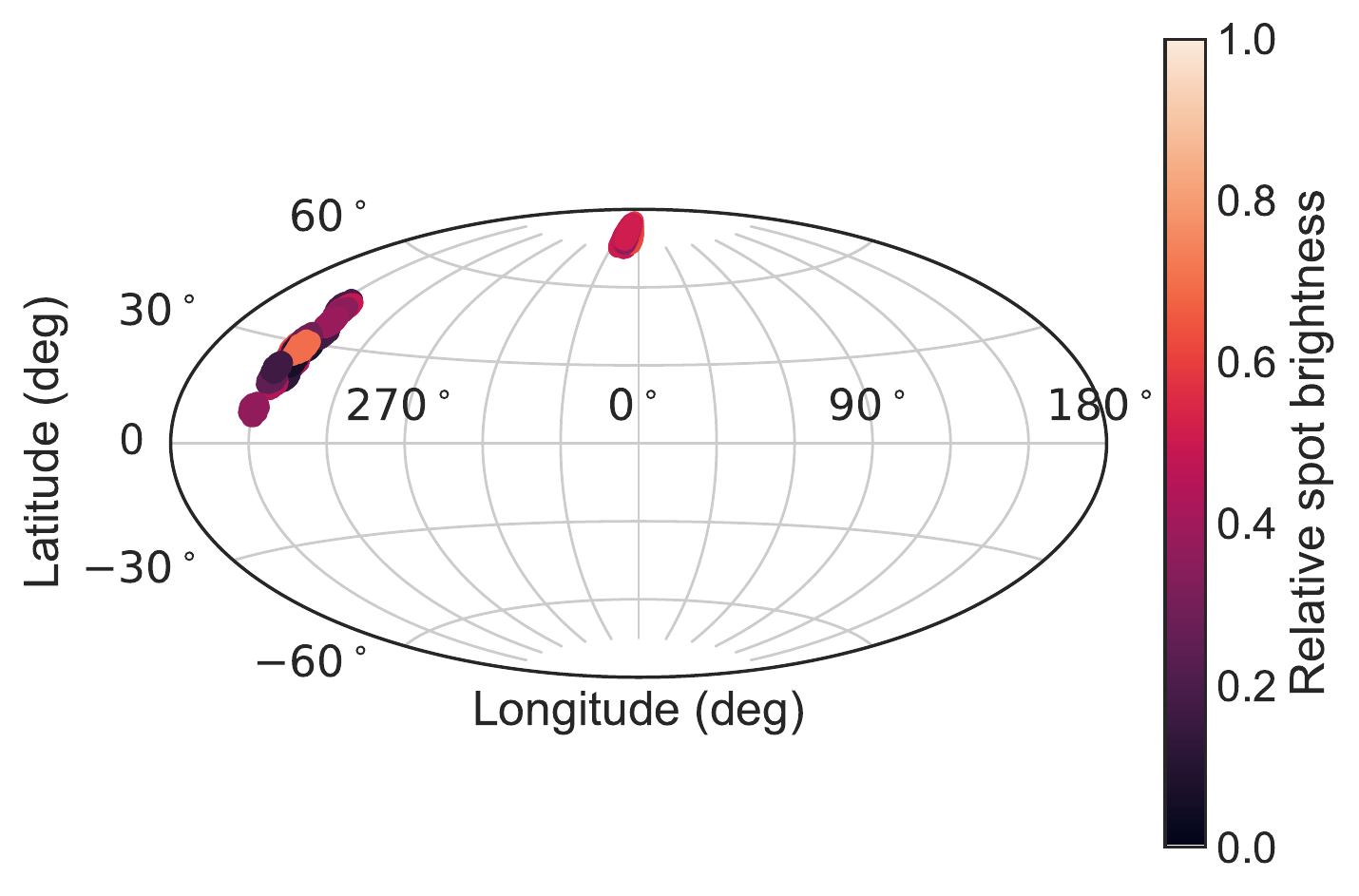}
    \caption{Spot maps for GJ 1243, with 20 randomly drawn posterior samples (i.e., 20 realizations of each spot) to show the range of possible models consistent with the data.
    The longitude and latitude are visualized in the Aitoff projection.
    Color coding represents the relative spot brightness compared to the stellar surface brightness in the \TESS{} band pass.}
    \label{fig:GJ_1243_spot_map}
\end{figure*}

We find that the model with two star spots and three flares describes the data best, according to the Bayes factors (Fig.~\ref{fig:GJ_1243_fit} and \ref{fig:GJ_1243_Bayes_factors}). In this model, the primary spot lies close to the pole (longitude $345.3\pm2.7$, latitude $79.5\pm1.6$) with a angular radius of ${3.51_{-0.98}^{+1.1}}^\circ$ and a relative brightness of $0.561_{-0.077}^{+0.060}$ compared to the stars surface brightness in the \TESS{} band (Fig.~\ref{fig:GJ_1243_spot_map}). This puts it at an effective temperature of about $2900$\,K. In comparison, the host stars temperature is about $3300$\,K \citep{Stassun2017}. The second spot lies slightly closer to the equator (longitude $215.1\pm1.6$, latitude $28.8_{-6.9}^{+7.6}$), and is smaller and darker (angular radius $5.44_{-0.39}^{+0.52}$, relative brightness $0.22_{-0.13}^{+0.14}$), corresponding to a spot temperature of about $2500$\,K.
The three flares we identify have amplitudes of 10\%, 4\% and 3\%, respectively, with the first two flares appearing in sequence and overlapping each other. These `outbursts' of multiple, subsequent flares are common on active M dwarfs, and can, to some extent, be disentangled into individual flares using Bayesian evidence \citep[][]{Guenther2020}, as also demonstrated here (Fig.~\ref{fig:GJ_1243_fit} and \ref{fig:GJ_1243_Bayes_factors}).

\section{Summary and Conclusion}
\label{s:Summary and Conclusion}

In this work, we introduced the \texttt{allesfitter} package to perform a global inference based on photometric and RV data. 
\texttt{allesfitter} unites various robust and well-tested generative models of exoplanets and stars to perform parameter inference and model testing. It provides a flexible graphical user interface as well as a \texttt{Python} API.
We illustrated a range of analyses to exemplify use cases, including multi-planet systems on eccentric orbits, transit timing variations, phase curves, eclipsing binaries, star spots, and stellar flares.
In all cases we found a good agreement with the original studies.

\acknowledgments
We acknowledge helpful discussions with Edward Gillen, N{\'e}stor Espinoza, Josh Speagle, Pierre Maxted, Daniel Foreman-Mackey, Ismael Mireles, Mariona Badenas-Agusti and the \TESS{} team during development of \texttt{allesfitter} and its various predecessors.
M.N.G. acknowledges support from MIT’s Kavli Institute as a Torres postdoctoral fellow. 
T.D. acknowledges support from MIT’s Kavli Institute as a Kavli postdoctoral fellow. 
\texttt{Allesfitter} is written in \texttt{python 3} \citep{Rossum1995}. 
In addition to the above mentioned model-specific packages, it also makes use of the open-source software
\texttt{corner} \citep{Foreman-Mackey2016},
\texttt{transitleastsquares} \citep{Hippke2019a},
\texttt{wotan} \citep{Hippke2019b},
\texttt{numpy} \citep{vanderWalt2011}, 
\texttt{scipy} \citep{Jones2001},
\texttt{matplotlib} \citep{Hunter2007}, 
\texttt{tqdm} (doi:10.5281/zenodo.1468033) and 
\texttt{seaborn} (\url{https://seaborn.pydata.org/index.html}).

\clearpage
\appendix
\renewcommand\thefigure{A\arabic{figure}}   
\setcounter{figure}{0} 
\renewcommand\thetable{A\arabic{table}}   
\setcounter{table}{0}

\input{tab_Settings} 

\input{tab_Parameters} 

\input{tab_Derived} 

\clearpage
\begin{figure*}[!htbp]
    \centering
    \includegraphics[width=\textwidth]{Figs/Pi_Mensae_ns_corner_compressed.pdf}
    \caption{Posteriors for the global fit to the two planet system Pi~Mensae, using all available data from \TESS{} Year 1 (Sectors 1, 4, 8, 11, 12 and 13) as well as all RV data used in \citet{Huang2018b}. Red lines are the published median values from \citet{Huang2018b}, which used \TESS{} Sector 1 data. The \texttt{allesfitter} posteriors agree well with the published values.}
    \label{fig:Pi_Mensae_ns_corner}
\end{figure*}

\clearpage
\begin{figure*}[!htbp]
    \centering
    \includegraphics[width=\textwidth]{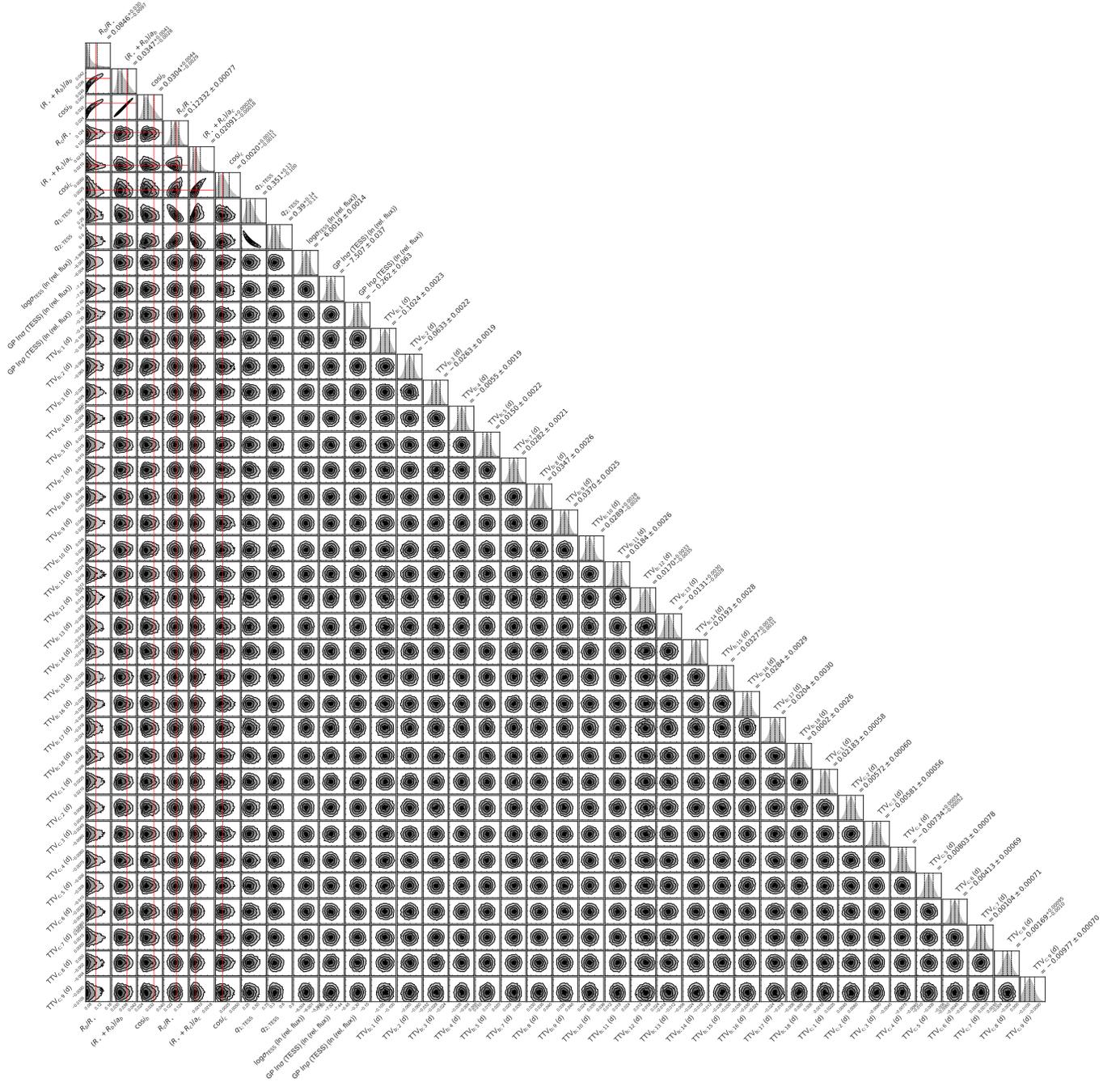}
    \caption{Posteriors for the global fit with free TTVs to the two planet system TOI-216, using all available data from \TESS{} Year 1 (Sectors 1-9 and 11-13). Red lines are the published median values from \citet{Dawson2019}, which used \TESS{} Sector 1-6 data. The \texttt{allesfitter} posteriors agree well with the published values.}
    \label{fig:TOI-216_ns_corner_compressed}
\end{figure*}

\clearpage
\begin{figure*}[!htbp]
    \centering
    \includegraphics[width=\textwidth]{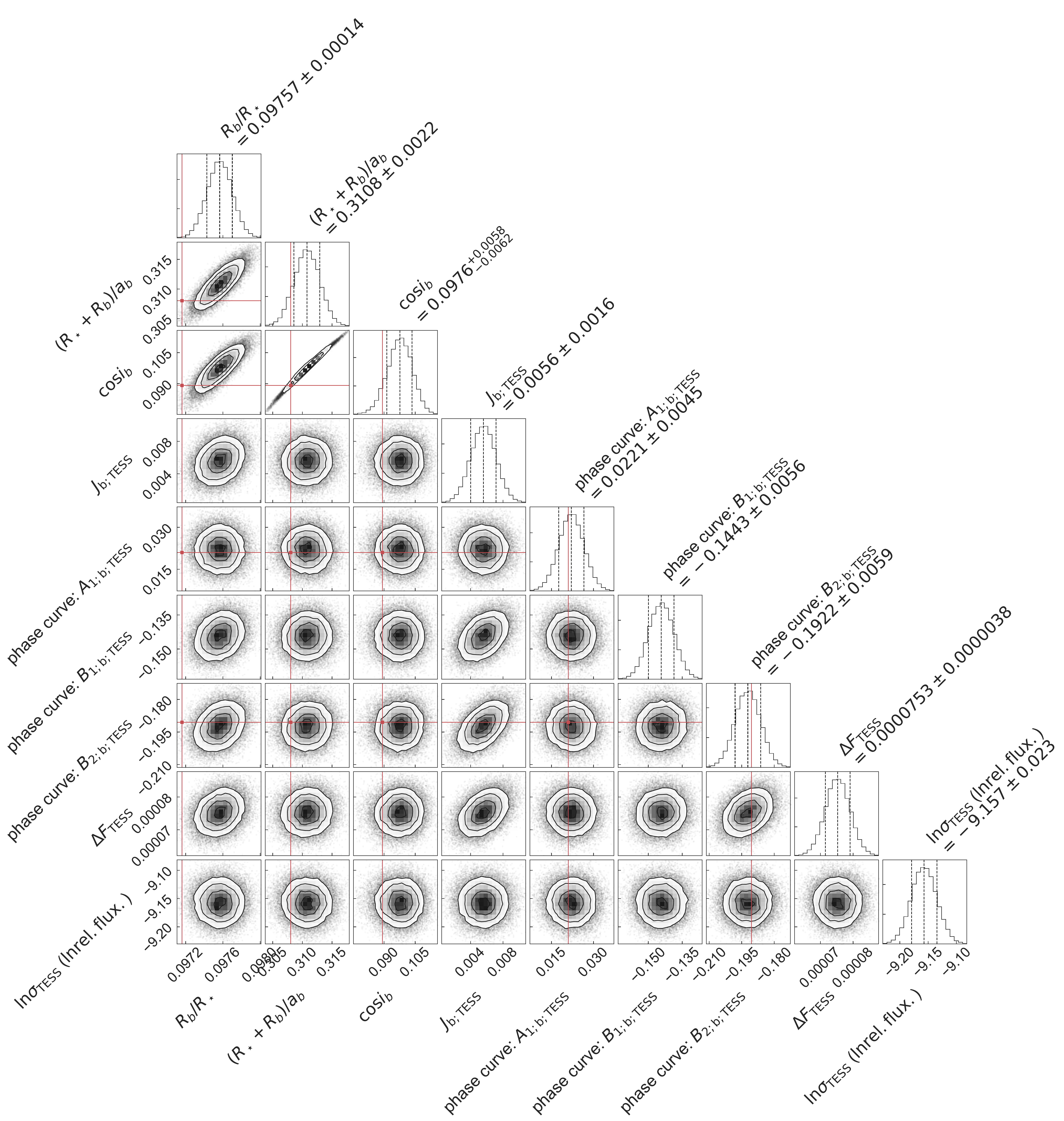}
    \caption{Posteriors for the global fit to the \TESS{} optical phase curve of WASP-18 using \texttt{allesfitter}'s sine\_series model. Red lines are the published values from \citep{Shporer2019}. 
    The \texttt{allesfitter} posteriors agree well with the published values overall.
    The deviations for the radius ratio and amplitude of the atmospheric modulation are likely due to our simplified example, which is run on a phase-folded and binned light curve with fixed parameters and constant baseline.}
    \label{fig:WASP-18_mcmc_corner}
\end{figure*}

\clearpage
\begin{figure*}[!htbp]
    \centering
    \includegraphics[width=\textwidth]{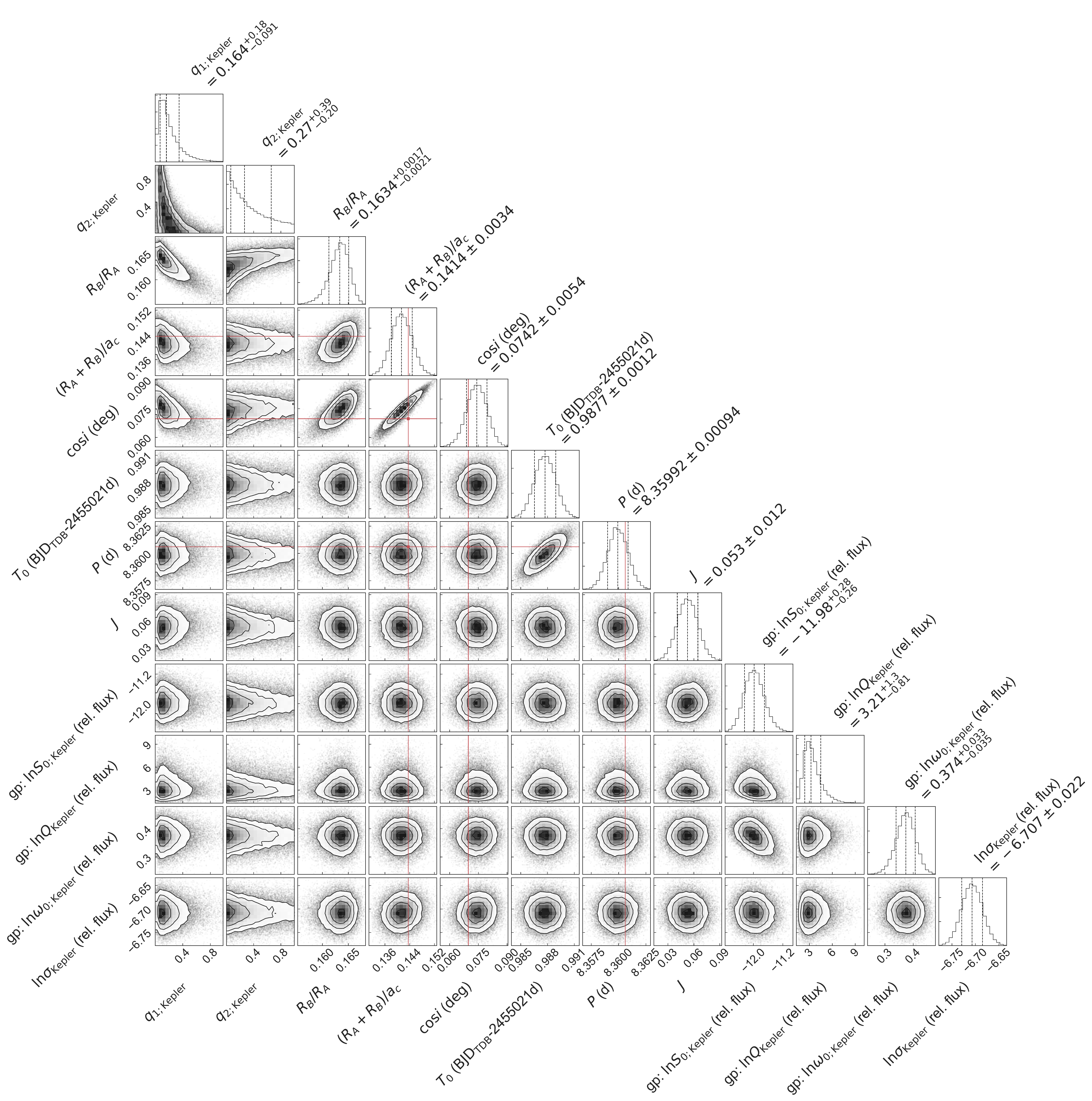}
    \caption{Posteriors for the global fit of the spotted binary star system KOI-1003, using the first 28\,days of the \Kepler{} Quarter 2 long-cadence light curve. Red lines are the published values from \citet{Roettenbacher2016}. The \texttt{allesfitter} posteriors agree well with the published values.
    The deviation for the radius ratio is likely caused by spot crossings in this section of the light curve, as also discussed by \citet{Roettenbacher2016}.}
    \label{fig:KOI-1003_mcmc_corner}
\end{figure*}

\clearpage
\begin{figure*}[!htbp]
    \centering
    \includegraphics[width=\textwidth]{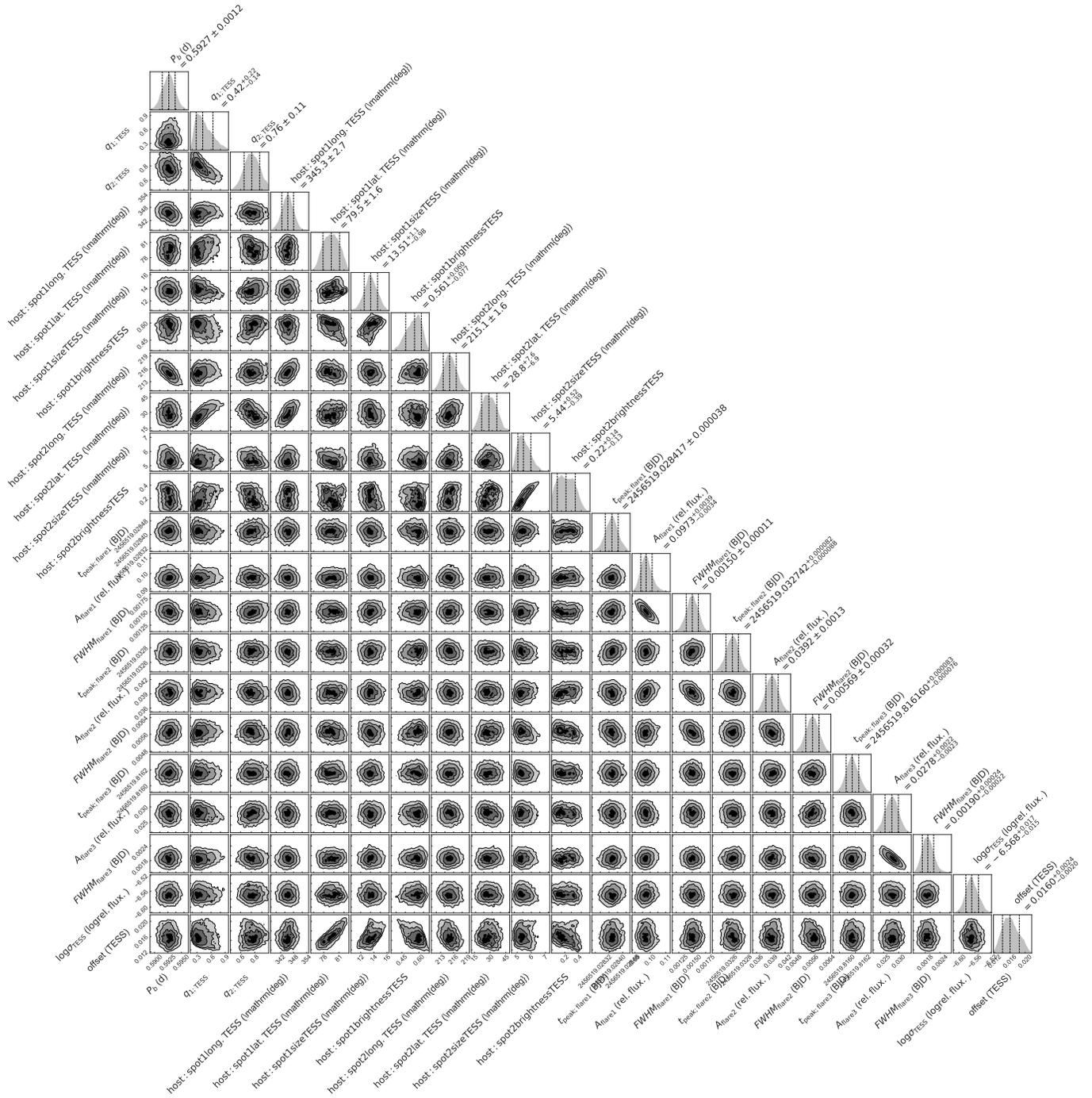}
    \caption{Posteriors for the global fit of GJ 1243, using a model of two spots and three flares for a 1.8\,day section of the \TESS{} short-cadence light curve.}
    \label{fig:GJ_1243_ns_corner}
\end{figure*}

\clearpage
\bibliography{references}

\end{document}

%% file: tab_Pi_Mensae_results.tex
\begin{table*}[!ht]
    \centering
	\caption{Updated parameters from the \texttt{allesfit} of Pi Mensae, using all available data from \TESS{} Year 1 (Sectors 1, 4, 8, 11, 12 and 13) as well as all RV data used in \citet{Huang2018b}.}
	\label{tab:Pi_Mensae_results}
	\centering
	\begin{tabular}{lll}
    \hline
    \hline
    Parameter & Value & Source \\ 
    \hline 
    \multicolumn{3}{c}{\textit{Fitted parameters}} \\ 
    \hline 
    Transformed limb darkening, $q_{1;\mathrm{TESS}}$ & $0.46_{-0.14}^{+0.18}$ & fit \\ 
    Transformed limb darkening, $q_{2;\mathrm{TESS}}$ & $0.21_{-0.15}^{+0.25}$ & fit \\ 
    Epoch b, $T_{0;b}$ ($\mathrm{BJD_{TDB}}$) & $2456552.4\pm2.5$ & fit \\ 
    Period b, $P_b$ (days) & $2093.1\pm1.8$ & fit \\ 
    RV semi-amplitude b, $K_b$ (km/s) & $0.1926\pm0.0013$ & fit \\ 
    Eccentricity term b, $\sqrt{e_b} \cos{\omega_b}$ & $0.6956\pm0.0043$ & fit \\ 
    Eccentricity term b, $\sqrt{e_b} \sin{\omega_b}$ & $-0.3919\pm0.0060$ & fit \\ 
    Sum of radii over semi-major axis c, $(R_\star + R_c) / a_c$ & $0.0761_{-0.0016}^{+0.0019}$ & fit \\ 
    Radius ratio c, $R_c / R_\star$ & $0.01696\pm0.00023$ & fit \\ 
    Cosine of inclination c, $\cos{i_c}$ & $0.0427_{-0.0031}^{+0.0033}$ & fit \\ 
    Epoch c, $T_{0;c}$ ($\mathrm{BJD_{TDB}}$) & $2458501.00304_{-0.00039}^{+0.00035}$ & fit \\ 
    Period c, $P_c$ (days) & $6.267850\pm0.000018$ & fit \\ 
    RV semi-amplitude c, $K_c$ (km/s) & $0.00153\pm0.00028$ & fit \\ 
    Eccentricity term c, $\sqrt{e_c} \cos{\omega_c}$ & $0.0$ & fixed \\ 
    Eccentricity term c, $\sqrt{e_c} \sin{\omega_c}$ & $0.0$ & fixed \\ 
    GP: $\ln \sigma_\mathrm{TESS}$ ($\ln$ rel. flux) & $-10.471\pm0.037$ & fit \\ 
    GP: $\ln \rho_\mathrm{TESS}$ ($\ln$ days) & $-1.98\pm0.11$ & fit \\ 
    RV offset, $\Delta RV_\mathrm{AAT}$ (km/s) & $0.03198\pm0.00085$ & fit \\ 
    RV offset, $\Delta RV_\mathrm{HARPS_1}$ (km/s) & $10.70848\pm0.00038$ & fit \\ 
    RV offset, $\Delta RV_\mathrm{HARPS_2}$ (km/s) & $10.73058\pm0.00069$ & fit \\ 
    Nat. log. error scaling, $\ln \sigma_\mathrm{TESS}$ ($\ln$ rel. flux) & $-8.6313\pm0.0024$ & fit \\ 
    Nat. log. jitter term, $\ln \sigma_\mathrm{AAT}$ ($\ln$ km/s) & $-5.013_{-0.087}^{+0.094}$ & fit \\ 
    Nat. log. jitter term, $\ln \sigma_\mathrm{HARPS_1}$ ($\ln$ km/s) & $-6.041\pm0.078$ & fit \\ 
    Nat. log. jitter term, $\ln \sigma_\mathrm{HARPS_2}$ ($\ln$ km/s) & $-6.40_{-0.18}^{+0.21}$ & fit \\ 
    \hline 
    \multicolumn{3}{c}{\textit{Derived parameters}} \\ 
    \hline 
    Eccentricity b, $e_\mathrm{b}$ & $0.6375\pm0.0024$ & derived \\ 
    Arg. of periastron b, $w_\mathrm{b}$ (deg) & $330.60\pm0.53$ & derived \\ 
    Period ratio, $P_\mathrm{b} / P_\mathrm{c}$ & $333.95\pm0.29$ & derived \\ 
    Host radius over semi-major axis c, $R_\star/a_\mathrm{c}$ & $0.0749_{-0.0016}^{+0.0019}$ & derived \\ 
    Semi-major axis c over host radius, $a_\mathrm{c}/R_\star$ & $13.36_{-0.33}^{+0.29}$ & derived \\ 
    Planet radius c over semi-major axis c, $R_\mathrm{c}/a_\mathrm{c}$ & $0.001270_{-0.000037}^{+0.000043}$ & derived \\ 
    Planet radius c, $R_\mathrm{c}$ ($\mathrm{R_{\oplus}}$) & $2.035\pm0.052$ & derived \\ 
    Semi-major axis c, $a_\mathrm{c}$ ($\mathrm{R_{\odot}}$) & $14.67\pm0.46$ & derived \\ 
    Semi-major axis c, $a_\mathrm{c}$ (AU) & $0.0682\pm0.0021$ & derived \\ 
    Inclination c, $i_\mathrm{c}$ (deg) & $87.55_{-0.19}^{+0.18}$ & derived \\ 
    Planet mass c, $M_\mathrm{c}$ ($\mathrm{M_{\oplus}}$) & $4.71_{-0.85}^{+0.90}$ & derived \\ 
    Impact parameter c, $b_\mathrm{tra;c}$ & $0.571\pm0.031$ & derived \\ 
    Total transit duration, $T_\mathrm{tot;c}$ (h) & $3.020_{-0.023}^{+0.028}$ & derived \\ 
    Full-transit duration, $T_\mathrm{full;c}$ (h) & $2.870_{-0.026}^{+0.031}$ & derived \\ 
    Stellar density from orbit c, $\rho_\mathrm{\star;c}$ (cgs) & $1.148_{-0.082}^{+0.077}$ & derived \\ 
    Planet density c, $\rho_\mathrm{c}$ (cgs) & $3.06_{-0.60}^{+0.67}$ & derived \\ 
    Planet surface gravity c, $g_\mathrm{\star;c}$ (cgs) & $1100_{-200}^{+210}$ & derived \\ 
    Equilibrium temperature c, $T_\mathrm{eq;c}$ (K) & $1069_{-14}^{+15}$ & derived \\ 
    Transit depth c, $\delta_\mathrm{tr; c; TESS}$ (ppt) & $0.3204_{-0.0066}^{+0.0083}$ & derived \\ 
    Limb darkening $u_\mathrm{1; TESS}$ & $0.28_{-0.20}^{+0.26}$ & derived \\ 
    Limb darkening $u_\mathrm{2; TESS}$ & $0.40_{-0.35}^{+0.28}$ & derived \\  
	\end{tabular}
\end{table*}

%% file: tab_TOI-216_results.tex
\begin{table*}[!ht]
    \centering
	\caption{Updated parameters from the \texttt{allesfit} of TOI-216, using all available data from \TESS{} Year 1 (Sectors 1-9 and 11-13).}
	\label{tab:TOI-216_results}
	\centering
	\begin{tabular}{lll}
    \hline
    \hline
    Parameter & Value & Source \\ 
    \hline 
    \multicolumn{3}{c}{\textit{Fitted parameters}} \\ 
    \hline 
    Radius ratio b, $R_b / R_\star$ & $0.0846_{-0.0097}^{+0.030}$ & fit \\ 
    Sum of radii over semi-major axis b, $(R_\star + R_b) / a_b$ & $0.0347_{-0.0028}^{+0.0041}$ & fit \\ 
    Cosine of inclination b, $\cos{i_b}$ & $0.0304_{-0.0029}^{+0.0044}$ & fit \\ 
    Linear-ephemerides epoch b, $T_{0;b}$ ($\mathrm{BJD_{TDB}}$) & $2458496.1366$ & fixed \\ 
    Linear-ephemerides period b, $P_b$ (days) & $17.0714$ & fixed \\ 
    Sum of radii over semi-major axis c, $R_c / R_\star$ & $0.12332\pm0.00077$ & fit \\ 
    Radius ratio c, $(R_\star + R_c) / a_c$ & $0.02091_{-0.00018}^{+0.00026}$ & fit \\ 
    Cosine of inclination c, $\cos{i_c}$ & $0.0020_{-0.0011}^{+0.0015}$ & fit \\ 
    Linear-ephemerides epoch c, $T_{0;c}$ ($\mathrm{BJD_{TDB}}$) & $2458504.0408$ & fixed \\ 
    Linear-ephemerides period c, $P_c$ (days) & $34.5555$ & fixed \\ 
    Transformed limb darkening, $q_{1; \mathrm{TESS}}$ & $0.351_{-0.100}^{+0.13}$ & fit \\ 
    Transformed limb darkening, $q_{2; \mathrm{TESS}}$ & $0.39_{-0.11}^{+0.14}$ & fit \\ 
    GP: $\ln \sigma_\mathrm{TESS}$ ($\ln$ rel. flux) & $-7.507\pm0.037$ & fit \\ 
    GP: $\ln \rho_\mathrm{TESS}$ ($\ln$ days) & $-0.262\pm0.063$ & fit \\ 
    Nat. log. error scaling, $\log{\sigma_\mathrm{TESS}}$ & $-6.0019\pm0.0014$ & fit \\ 
    \hline 
    \multicolumn{3}{c}{\textit{Derived parameters}} \\ 
    \hline 
    Host radius over semi-major axis b, $R_\star/a_\mathrm{b}$ & $0.0320_{-0.0023}^{+0.0028}$ & derived \\ 
    Semi-major axis b over host radius, $a_\mathrm{b}/R_\star$ & $31.3\pm2.5$ & derived \\ 
    Planet radius b over semi-major axis b, $R_\mathrm{b}/a_\mathrm{b}$ & $0.00270_{-0.00047}^{+0.0013}$ & derived \\ 
    Planet radius b, $R_\mathrm{b}$ ($\mathrm{R_{\oplus}}$) & $3.53_{-0.45}^{+1.3}$ & derived \\ 
    Semi-major axis b, $a_\mathrm{b}$ ($\mathrm{R_{\odot}}$) & $11.9_{-1.1}^{+1.2}$ & derived \\ 
    Semi-major axis b, $a_\mathrm{b}$ (AU) & $0.0552_{-0.0050}^{+0.0054}$ & derived \\ 
    Inclination b, $i_\mathrm{b}$ (deg) & $88.26_{-0.25}^{+0.17}$ & derived \\ 
    Impact parameter b, $b_\mathrm{tra;b}$ & $0.952_{-0.026}^{+0.051}$ & derived \\ 
    Total transit duration b, $T_\mathrm{tot;b}$ (h) & $2.163\pm0.068$ & derived \\ 
    Stellar density from orbit b, $rho_\mathrm{\star;b}$ (cgs) & $1.99_{-0.44}^{+0.51}$ & derived \\ 
    Equilibrium temperature b, $T_\mathrm{eq;b}$ (K) & $392\pm22$ & derived \\ 
    Transit depth b, $\delta_\mathrm{tr; b; TESS}$ (ppt) & $0.00448_{-0.00017}^{+0.00014}$ & derived \\ 
    Host radius over semi-major axis c, $R_\star/a_\mathrm{c}$ & $0.01862_{-0.00016}^{+0.00023}$ & derived \\ 
    Semi-major axis c over host radius, $a_\mathrm{c}/R_\star$ & $53.72_{-0.66}^{+0.47}$ & derived \\ 
    Planet radius c over semi-major axis c, $R_\mathrm{c}/a_\mathrm{c}$ & $0.002291_{-0.000018}^{+0.000036}$ & derived \\ 
    Planet radius c, $R_\mathrm{c}$ ($\mathrm{R_{\oplus}}$) & $5.11\pm0.27$ & derived \\ 
    Semi-major axis c, $a_\mathrm{c}$ ($\mathrm{R_{\odot}}$) & $20.4\pm1.1$ & derived \\ 
    Semi-major axis c, $a_\mathrm{c}$ (AU) & $0.0948\pm0.0052$ & derived \\ 
    Inclination c, $i_\mathrm{c}$ (deg) & $89.883_{-0.089}^{+0.066}$ & derived \\ 
    Impact parameter c, $b_\mathrm{tra;c}$ & $0.110_{-0.062}^{+0.081}$ & derived \\ 
    Total transit duration c, $T_\mathrm{tot;c}$ (h) & $5.487_{-0.032}^{+0.036}$ & derived \\ 
    Full transit duration c, $T_\mathrm{full;c}$ (h) & $4.262_{-0.033}^{+0.036}$ & derived \\ 
    Stellar density from orbit c, $rho_\mathrm{\star;c}$ (cgs) & $2.456_{-0.089}^{+0.066}$ & derived \\ 
    Equilibrium temperature c, $T_\mathrm{eq;c}$ (K) & $299\pm12$ & derived \\ 
    Transit depth c, $\delta_\mathrm{tr; c; TESS}$ (ppt) & $0.01829_{-0.00013}^{+0.00014}$ & derived \\ 
    Limb darkening $u_\mathrm{1; TESS}$ & $0.466\pm0.075$ & derived \\ 
    Limb darkening $u_\mathrm{2; TESS}$ & $0.12_{-0.16}^{+0.17}$ & derived \\ 
    Median stellar density from orbits, $rho_\mathrm{\star}$ (cgs) & $2.39_{-0.63}^{+0.13}$ & derived \\ 
	\end{tabular}
\end{table*}

%% file: tab_TOI-216_o_minus_c.tex
\begin{table}[!ht]
    \centering
	\caption{{\revision Updated transit mid-times and TTV O-C values for TOI-216 from all \TESS{} Year 1 data (Sectors 1-9 and 11-13).}}
    \label{tab:TOI-216_o_minus_c}
	\centering
	\begin{tabular}{cc}
    \hline
    \hline
    Transit mid-time & O-C\\
    ($\mathrm{BJD_{TDB}}$) & (min.)\\
    \hline
    \multicolumn{2}{c}{TOI-216 b}\\
    \hline
    $2458325.3202\pm0.0023$ & $-0.0857\pm0.0023$\\
    $2458342.4307\pm0.0022$ & $-0.0476\pm0.0022$\\
    $2458359.5391\pm0.0019$ & $-0.0115\pm0.0019$\\
    $2458376.6313\pm0.0019$ & $0.0084\pm0.0019$\\
    $2458393.7232\pm0.0022$ & $0.0280\pm0.0022$\\
    $2458427.8792\pm0.0021$ & $0.0392\pm0.0021$\\
    $2458444.9571\pm0.0026$ & $0.0448\pm0.0026$\\
    $2458462.0308\pm0.0025$ & $0.0462\pm0.0025$\\
    $2458479.0941_{-0.0026}^{+0.0028}$ & $0.0372_{-0.0026}^{+0.0028}$\\
    $2458496.1550\pm0.0026$ & $0.0257\pm0.0026$\\
    $2458513.2250_{-0.0035}^{+0.0033}$ & $0.0234_{-0.0035}^{+0.0033}$\\
    $2458547.3377_{-0.0028}^{+0.0030}$ & $-0.0086_{-0.0028}^{+0.0030}$\\
    $2458564.4029\pm0.0028$ & $-0.0157\pm0.0028$\\
    $2458615.6037_{-0.0031}^{+0.0030}$ & $-0.0319_{-0.0031}^{+0.0030}$\\
    $2458632.6794\pm0.0029$ & $-0.0286\pm0.0029$\\
    $2458649.7588\pm0.0030$ & $-0.0215\pm0.0030$\\
    $2458666.8508\pm0.0026$ & $-0.0018\pm0.0026$\\
    \hline
    \multicolumn{2}{c}{TOI-216 c}\\
    \hline
    $2458331.28513\pm0.00058$ & $0.01593\pm0.00058$\\
    $2458365.82452\pm0.00060$ & $0.00125\pm0.00060$\\
    $2458400.36849\pm0.00056$ & $-0.00886\pm0.00056$\\
    $2458434.92246_{-0.00052}^{+0.00054}$ & $-0.00897_{-0.00052}^{+0.00054}$\\
    $2458469.47727\pm0.00078$ & $-0.00822\pm0.00078$\\
    $2458538.59217\pm0.00069$ & $-0.00148\pm0.00069$\\
    $2458607.70834\pm0.00071$ & $0.00654\pm0.00071$\\
    $2458642.26111_{-0.0010}^{+0.00095}$ & $0.00523_{-0.0010}^{+0.00095}$\\
    $2458676.80853\pm0.00070$ & $-0.00142\pm0.00070$\\
    \hline
	\end{tabular}
\end{table}

%% file: tab_Settings.tex
\clearpage
\begin{table*}[ht]
    \centering
    
	\caption{All possible settings for \texttt{allesfitter}, which can be given in the \textit{settings.csv} file. This list reflects \texttt{allesfitter} version 1.1.1. For future additions and the most up to date documentation see \url{www.allesfitter.com}.}
    \label{tab:Settings}
	\centering
	\begin{tabular}{p{4cm}p{11cm}p{2cm}}
    \hline
    \hline
    Setting & Explanation & Default\\
    \hline
    
    &&\\
    \multicolumn{3}{c}{General settings} \\
    \hline
    \text{companions\_phot} & The companion(s) in the photometric data, space separated & - \\
     & Example: \textit{companions\_phot,b c e} & \\
    \text{companions\_rv} & The companion(s) in the radial velocity data, space separated & - \\ 
     & Example: \textit{companions\_rv,B} & \\
    \text{inst\_phot} & The photometric instrument(s), space separated & - \\
     & Example: \textit{inst\_phot,TESS} & \\
    \text{inst\_rv} & The radial velocity instrument(s), space separated & - \\
     & Example: \textit{inst\_rv,HARPS ESPRESSO} & \\
    
    &&\\
    \multicolumn{3}{c}{Fit performance settings} \\
    \hline
    \text{multiprocess} & Use multiprocessing (\textit{True/False}) & False\\
    \text{multiprocess\_cores} & Number of cores for multiprocessing (\textit{1,2,3...,all}) & 1\\
    \text{fast\_fit} & Mask out the out-of-transit data (\textit{True/False}) & False\\
    \text{fast\_fit\_width} & If using fast fit, select the window size around the transit (in days) & 0.33333\\
    \text{secondary\_eclipse} & If using fast fit, also keep a window around phase 0.5 (\textit{True/False}) & False\\
    \text{phase\_curve} & Generate output and figures for phase curves (\textit{True/False}) & False\\
    \text{phase\_curve\_style} & Which phase curve model to use (see Section~\ref{ss:Phase curves}; \textit{None\slash sine\_physical\slash \linebreak sine\_series\slash ellc\_physical}) & None\\
    \text{shift\_epoch} & Shift the epoch into the middle of the data set (\textit{True/False}) & False\\
    \text{inst\_for\_[comp]\_epoch} & If using shift epoch, which data files should be used & - \\
    
    &&\\
    \multicolumn{3}{c}{MCMC settings} \\
    \hline
    \text{mcmc\_nwalkers} & Number of MCMC walkers & 100\\
    \text{mcmc\_total\_steps} &  Total steps in the MCMC chain, including burn-in steps & 2000\\
    \text{mcmc\_burn\_steps} & Burn-in steps in the MCMC chain & 1000\\
    \text{mcmc\_thin\_by} & Only save every \textit{n}-th step in the MCMC chain & 1\\
    \text{mcmc\_pre\_run\_steps} & Run \textit{n} steps of pre-burn-in to refine the initial guess & 0\\
    \text{mcmc\_pre\_run\_loops} & Run \textit{m} loops with the above \textit{n} steps of pre-burn-ins & 0\\

    &&\\
    \multicolumn{3}{c}{Nested Sampling settings} \\
    \hline
    \text{ns\_modus} & Nested Sampling algorithm (\textit{static/dynamic}) & dynamic\\
    \text{ns\_nlive} & Number of live points & 500\\
    \text{ns\_bound} & Method to bound the prior (\textit{None/single/multi/balls/cubes}) & single\\
    \text{ns\_sample} & Method to update live points (\textit{auto/unif/rwalk/rstagger/slice/rslice\slash hslice}) & rwalk\\
    \text{ns\_tol} & Tolerance of the convergence criterion & 0.01\\

    &&\\
    \multicolumn{3}{c}{External priors: stellar host density} \\
    \hline
    \text{use\_host\_density\_prior} & Use an external normal prior on the host density (see Section~\ref{ss:External priors: stellar host density}; \textit{True/False}) & True\\
    
    &&\\
    \multicolumn{3}{c}{Limb darkening law per object and instrument} \\
    \hline
    \text{host\_ld\_law\_[inst]} & Limb darkening law for the host (see Section~\ref{ss:Limb darkening parameterizations}; none/lin/quad/sing) & none\\
    \text{[comp]\_ld\_law\_[inst]} & Limb darkening law for a companion (see Section~\ref{ss:Limb darkening parameterizations}; \textit{none/lin/quad\slash sing}) & none\\
    
    &&\\
	\multicolumn{3}{r}{... continued on next page ...}
	\end{tabular}
\end{table*}

\addtocounter{table}{-1}
\begin{table*}[ht]
    \centering
	\caption{... continued from previous page ...}
	\centering
	\begin{tabular}{p{4cm}p{11cm}p{2cm}}
    \hline
    \hline
    Setting & Explanation & Default\\    
    \hline
    
    &&\\
    \multicolumn{3}{c}{Baseline settings per instrument} \\
    \hline
    \text{baseline\_[key]\_[inst]} & The baseline method used per instrument (see Section~\ref{ss:Baselines (red noise)}, 
    \textit{none\slash 
    sample\_offset\slash 
    sample\_linear\slash 
    sample\_GP\_real\slash 
    sample\_GP\_complex\slash 
    sample\_GP\_Matern32\slash 
    sample\_GP\_SHO\slash 
    hybrid\_offset\slash 
    hybrid\_poly\_1\slash 
    hybrid\_poly\_2\slash 
    hybrid\_poly\_3\slash 
    hybrid\_poly\_4\slash 
    hybrid\_spline}) & none\\
    
    &&\\
    \multicolumn{3}{c}{Error settings per instrument} \\
    \hline
    \text{error\_[key]\_[inst]} & The white noise method per instrument, which either scales the noise for photometry, or adds a jitter term in quadrature for RV data (see Section~\ref{ss:White noise and jitter terms}; \textit{sample\slash hybrid}) & sample\\
    
    &&\\
    \multicolumn{3}{c}{Exposure interpolation} \\
    \hline
    \text{t\_exp\_[inst]} & Exposure time of the instrument (in days); crucial for long exposures or binned data, to sample a high cadence light curve model and bin it up to match the data binning & None\\
     & Example for 30 min cadence: \textit{t\_exp\_[inst],0.0208333} & \\
    \text{t\_exp\_n\_int\_[inst]} & Number of fine sampling points for the exposure interpolation & None\\
     & Example for 30 min cadence: \textit{t\_exp\_n\_int\_[inst],10} & \\
    
    &&\\
    \multicolumn{3}{c}{Stellar spots per object and instrument} \\
    \hline
    \text{host\_N\_spots\_[inst]} & Number of star spots on the host to include in the model; this will unlock the respective rows in the parameters file & 0\\
    \text{[comp]\_N\_spots\_[inst]} & Number of star spots on the companion to include in the model; this will unlock the respective rows in the parameters file & 0\\
    
    &&\\
    \multicolumn{3}{c}{Stellar flares} \\
    \hline
    \text{N\_flares} & Number of stellar flares to include in the model; this will unlock the respective rows in the parameters file & 0\\
    
    &&\\
    \multicolumn{3}{c}{Transit timing variations} \\
    \hline
    \text{fit\_ttvs} & Address transit/eclipse timing variations by freely fitting each transit/eclipse midtime; this will unlock the respective rows in the parameters file & False \\
    
    &&\\
    \multicolumn{3}{c}{Stellar grid per object and instrument} \\
    \hline
    \text{host\_grid\_[inst]} & How finely to integrate over the surface of the host star (see Section~\ref{ss:Stellar/planetary grid and shape}; \textit{very\_sparse/sparse/default/fine/very\_fine}) & default \\
    \text{[comp]\_grid\_[inst]} & How finely to integrate over the surface of the companion (see Section~\ref{ss:Stellar/planetary grid and shape}; \textit{very\_sparse/sparse/default/fine/very\_fine}) & default \\
    
    &&\\
    \multicolumn{3}{c}{Stellar shape per object and instrument} \\
    \hline
    \text{host\_shape\_[inst]} & How to compute the shape of the host star (see Section~\ref{ss:Stellar/planetary grid and shape}; 
    \textit{sphere\slash 
            roche\slash 
            roche\_v\slash 
            poly1p5\slash 
            poly3p0\slash love}) & default \\
    \text{[comp]\_shape\_[inst]} & How to compute the shape of the companion (see Section~\ref{ss:Stellar/planetary grid and shape};
    \textit{sphere\slash 
            roche\slash 
            roche\_v\slash 
            poly1p5\slash 
            poly3p0\slash love}) & default \\
            
    &&\\
    \multicolumn{3}{c}{Flux weighted RVs per object and instrument} \\
    \hline
    \text{flux\_weighted} & Compute the flux-weighted RV over the objects' entire surface (e.g. for Rossiter-McLaughlin effect) or the RV of their center-of-mass (\textit{True/False}) & False \\
    
	\vspace{11pt}
    \end{tabular}
    \begin{minipage}{\textwidth} 
    \text{[comp]}: placeholder for the actual name given to the companion\\
    \text{[inst]}: placeholder for the actual name given to the instrument\\
    \text{[key]}: placeholder for either \textit{flux} or \textit{rv}
    \end{minipage}
\end{table*}

%% file: tab_Parameters.tex
\clearpage
\begin{table*}[ht]
    \centering
	\caption{A list of all possible parameters for \texttt{allesfitter}, which can be given in the \textit{params.csv} file. Note that not all of these can be selected at the same time, as some combinations depend on which models are chosen (for example, either a linear or a quadratic limb darkening model). This list reflects \texttt{allesfitter} version 1.1.1. For future additions and the most up to date documentation see \url{www.allesfitter.com}.}
    \label{tab:Parameters}
	\centering
	\begin{tabular}{lp{10cm}l}
    \hline
    \hline
    Parameter & Explanation & Default\\
    \hline
    &&\\
    \multicolumn{3}{c}{Frequently used astrophysical parameters} \\
    \hline
    \text{[comp]\_rr} & The radius ratio of companion to host, $R_\mathrm{comp} / R_\star$ & 0 \\
    \text{[comp]\_rsuma} & The sum of stellar and companion radii divided by the semi-major axis, $(R_\mathrm{comp}+{R_\star}) / a$  & 0 \\
    \text{[comp]\_cosi} & The cosine of the orbit of this companion, $\cos{i}$ & 0 \\
    \text{[comp]\_epoch} & The epoch / transit midtime in days, $T_0$ & 0 \\
    \text{[comp]\_period} & The orbital period of the companion in days, $P$ & 0 \\
    \text{[comp]\_K} & The host's RV semi-amplitude caused by the companio, $K$ & 0 \\
    \text{[comp]\_f\_c} & Transformation of eccentricity and argument of periastron, $\sqrt{e} \cos{\omega}$  & 0 \\
    \text{[comp]\_f\_s} & Transformation of eccentricity and argument of periastron, $\sqrt{e} \sin{\omega}$  & 0 \\
    \text{[comp]\_sbratio\_[inst]} & Surface brightness ratio between the companion and host star, $J$ & 0 \\
    \text{dil\_[inst]} & Dilution of the signal in the given instruments bandpass, & 0 \\
     & $D_0 := 1 - (F_\mathrm{source} / (F_\mathrm{source} + F_\mathrm{blend}))$. & \\
    
    &&\\
    \multicolumn{3}{c}{Limb darkening coefficients - linear/quadratic/three-parameter law (see Section~\ref{ss:Limb darkening parameterizations})} \\
    \hline
    \text{host\_ldc\_q1\_[inst]} & Transformed coefficient $q_1$ for host (lin/quad./ three-param.)& 0\\
    \text{host\_ldc\_q2\_[inst]} & Transf. coeff. $q_2$ for host (quad./three-param.)& 0\\
    \text{host\_ldc\_q3\_[inst]} & Transf. coeff. $q_3$ for host (three-param.)& 0\\
    
    \text{[comp]\_ldc\_q1\_[inst]} & Transf. coeff. $q_1$ for companion (lin/quad./ three-param.) & 0\\
    \text{[comp]\_ldc\_q2\_[inst]} & Transf. coeff. $q_2$ for companion (quad./ three-param.)& 0\\
    \text{[comp]\_ldc\_q3\_[inst]} & Transf. coeff. $q_3$ for companion (three-param.)& 0\\

    &&\\
    \multicolumn{3}{c}{Errors (white noise) (see Section~\ref{ss:White noise and jitter terms})} \\
    \hline
    \text{ln\_err\_flux\_[inst]} & Natural logarithm of error scaling for photometry, gets multiplied with the weights for the user-given errors & 0\\
    \text{ln\_jitter\_rv\_[inst]} & Natural logarithm of jitter term for RV, gets added in quadrature to the user-given errors & 0\\

    &&\\
    \multicolumn{3}{c}{Baselines (red noise) - constant offset (see Section~\ref{ss:Baselines (red noise)})} \\
    \hline
    \text{baseline\_offset\_[key]\_[inst]} & Constant offset & 0\\
    
    &&\\
    \multicolumn{3}{c}{Baselines (red noise) - linear trend (see Section~\ref{ss:Baselines (red noise)})} \\
    \hline
    \text{baseline\_offset\_[key]\_[inst]} & Constant offset & 0\\
    \text{baseline\_slope\_[key]\_[inst]} & Linear slope & 0\\   
    
    &&\\
    \multicolumn{3}{c}{Baselines (red noise) - GP with real kernel (see Section~\ref{ss:Baselines (red noise)})} \\
    \hline
    \text{baseline\_gp\_offset\_[key]\_[inst]} & Constant offset (optional; default 1 for flux, 0 for RV) & 0\\
    \text{baseline\_gp\_real\_lna\_[key]\_[inst]} & Natural logarithm of $a$ & 0\\
    \text{baseline\_gp\_real\_lnc\_[key]\_[inst]} & Natural logarithm of $c$ & 0\\   
    
    &&\\
	\multicolumn{3}{r}{... continued on next page ...}
	\end{tabular}
\end{table*}

\addtocounter{table}{-1}
\begin{table*}[ht]
    \centering
	\caption{... continued from previous page ...}
	\centering
	\begin{tabular}{lp{10cm}l}
    \hline
    \hline
    Parameter & Explanation & Default\\    
    \hline
    
    &&\\
    \multicolumn{3}{c}{Baselines (red noise) - GP with complex kernel (see Section~\ref{ss:Baselines (red noise)})} \\
    \hline
    \text{baseline\_gp\_offset\_[key]\_[inst]} & Constant offset (optional; default 1 for flux, 0 for RV) & 0\\
    \text{baseline\_gp\_complex\_lna\_[key]\_[inst]} & Natural logarithm of $a$ & 0\\
    \text{baseline\_gp\_complex\_lnb\_[key]\_[inst]} & Natural logarithm of $b$ & 0\\   
    \text{baseline\_gp\_complex\_lnc\_[key]\_[inst]} & Natural logarithm of $c$ & 0\\
    \text{baseline\_gp\_complex\_lnd\_[key]\_[inst]} & Natural logarithm of $d$ & 0\\   
    
    &&\\
    \multicolumn{3}{c}{Baselines (red noise) - GP with Mat{\'e}rn 3/2 kernel (see Section~\ref{ss:Baselines (red noise)})} \\
    \hline
    \text{baseline\_gp\_offset\_[key]\_[inst]} & Constant offset (optional; default 1 for flux, 0 for RV) & 0\\
    \text{baseline\_gp\_matern32\_lnsigma\_[key]\_[inst]} & Natural logarithm of $\sigma$ & 0\\
    \text{baseline\_gp\_matern32\_lnrho\_[key]\_[inst]} & Natural logarithm of $\rho$ & 0\\
    
    &&\\
    \multicolumn{3}{c}{Baselines (red noise) - GP with SHO kernel (see Section~\ref{ss:Baselines (red noise)})} \\
    \hline
    \text{baseline\_gp\_offset\_[key]\_[inst]} & Constant offset (optional; default 1 for flux, 0 for RV) & 0\\
    \text{baseline\_gp\_sho\_lnS0\_[key]\_[inst]} & Natural logarithm of $S_0$ & 0\\
    \text{baseline\_gp\_sho\_lnQ\_[key]\_[inst]} & Natural logarithm of $Q$ & 0\\
    \text{baseline\_gp\_sho\_lnomega0\_[key]\_[inst]} & Natural logarithm of $\omega_0$ & 0\\
    
    &&\\
    \multicolumn{3}{c}{Stellar variability - linear trend (see Section~\ref{ss:Stellar variability})} \\
    \hline
    \text{stellar\_var\_gp\_offset\_[key]\_[inst]} & Constant offset (optional; default 1 for flux, 0 for RV) & 0\\
    \text{stellar\_var\_offset\_[key]\_[inst]} & Constant offset & 0\\
    \text{stellar\_var\_slope\_[key]\_[inst]} & Linear slope & 0\\   
    
    &&\\
    \multicolumn{3}{c}{Stellar variability - GP with real kernel (see Section~\ref{ss:Stellar variability})} \\
    \hline
    \text{stellar\_var\_gp\_offset\_[key]\_[inst]} & Constant offset (optional; default 1 for flux, 0 for RV) & 0\\
    \text{stellar\_var\_gp\_real\_lna\_[key]\_[inst]} & Natural logarithm of $a$ & 0\\
    \text{stellar\_var\_gp\_real\_lnc\_[key]\_[inst]} & Natural logarithm of $c$ & 0\\   
    
    &&\\
    \multicolumn{3}{c}{Stellar variability - GP with complex kernel (see Section~\ref{ss:Stellar variability})} \\
    \hline
    \text{stellar\_var\_gp\_offset\_[key]\_[inst]} & Constant offset (optional; default 1 for flux, 0 for RV) & 0\\
    \text{stellar\_var\_gp\_complex\_lna\_[key]\_[inst]} & Natural logarithm of $a$ & 0\\
    \text{stellar\_var\_gp\_complex\_lnb\_[key]\_[inst]} & Natural logarithm of $b$ & 0\\   
    \text{stellar\_var\_gp\_complex\_lnc\_[key]\_[inst]} & Natural logarithm of $c$ & 0\\
    \text{stellar\_var\_gp\_complex\_lnd\_[key]\_[inst]} & Natural logarithm of $d$ & 0\\   
    
    &&\\
    \multicolumn{3}{c}{Stellar variability - GP with Mat{\'e}rn 3/2 kernel (see Section~\ref{ss:Stellar variability})} \\
    \hline
    \text{stellar\_var\_gp\_offset\_[key]\_[inst]} & Constant offset (optional; default 1 for flux, 0 for RV) & 0\\
    \text{stellar\_var\_gp\_matern32\_lnsigma\_[key]\_[inst]} & Natural logarithm of $\sigma$ & 0\\
    \text{stellar\_var\_gp\_matern32\_lnrho\_[key]\_[inst]} & Natural logarithm of $\rho$ & 0\\
    
    &&\\
    \multicolumn{3}{c}{Stellar variability - GP with SHO kernel (see Section~\ref{ss:Stellar variability})} \\
    \hline
    \text{stellar\_var\_gp\_sho\_lnS0\_[key]\_[inst]} & Natural logarithm of $S_0$ & 0\\
    \text{stellar\_var\_gp\_sho\_lnQ\_[key]\_[inst]} & Natural logarithm of $Q$ & 0\\
    \text{stellar\_var\_gp\_sho\_lnomega0\_[key]\_[inst]} & Natural logarithm of $\omega_0$ & 0\\
    
    &&\\
	\multicolumn{3}{r}{... continued on next page ...}
	\end{tabular}
\end{table*}

\addtocounter{table}{-1}
\begin{table*}[ht]
    \centering
	\caption{... continued from previous page ...}
	\centering
	\begin{tabular}{lp{9cm}l}
    \hline
    \hline
    Parameter & Explanation & Default\\    
    \hline
    &&\\
    \multicolumn{3}{c}{Phase curve parameters - sine\_series model (see Section~\ref{sss:Phase curves using sines})} \\
    \hline
    &&\\
    \multicolumn{3}{c}{Standard set:} \\
    \text{[comp]\_phase\_curve\_A1\_[inst]} & Semi-amplitude of the sine term $A_1 \sin{\Phi(t)}$ which approximates the Doppler boosting (beaming) modulation  & None \\
    \text{[comp]\_phase\_curve\_B1\_[inst]} & Semi-amplitude of the cosine term $B_1 \cos{\Phi(t)}$ which approximates the atmospheric (thermal and reflected light) modulation  & None \\
    \text{[comp]\_phase\_curve\_B1\_shift\_[inst]} & Time shift $s$ of the cosine term $B_1 \cos{\Phi(t + s)}$ (in days) & 0 \\
    \text{[comp]\_phase\_curve\_B2\_[inst]} & Semi-amplitude of the cosine term $B_2 \cos{2\Phi(t)}$ which approximates the leading-order  ellipsoidal (tidal distortion) modulation  & None \\
    \text{[comp]\_phase\_curve\_B3\_[inst]} & Semi-amplitude of the cosine term $B_3 \cos{3\Phi(t)}$ which approximates the next-order ellipsoidal (tidal distortion) modulation; this is usually negligible for exoplanets but can become measurable for binary stars & None \\
    
    &&\\
    \multicolumn{3}{c}{For differentiating thermal emission and reflected light one can use:} \\
    \text{[comp]\_phase\_curve\_B1t\_[inst]} & Semi-amplitude of the cosine term $B_{1t} \cos{\Phi(t)}$ which approximates the thermal emission part of the atmospheric modulation & None \\
    \text{[comp]\_phase\_curve\_B1t\_shift\_[inst]} & Time shift $s$ of the cosine term $B_{1t} \cos{\Phi(t + s)}$ (in days) & 0 \\
    \text{[comp]\_phase\_curve\_B1r\_[inst]} & Semi-amplitude of the cosine term $B_{1r} \cos{\Phi(t)}$ which approximates the reflected light part of the atmospheric modulation & None \\
    \text{[comp]\_phase\_curve\_B1r\_shift\_[inst]} & Time shift $s$ of the cosine term $B_{1r} \cos{\Phi(t + s)}$ (in days) & 0 \\
    
    &&\\
    \multicolumn{3}{c}{Phase curve parameters - sine\_physical model (see Section~\ref{sss:Phase curves using transformed sines})} \\
    \hline
    &&\\
    \multicolumn{3}{c}{Standard set:} \\
    
    \text{[comp]\_phase\_curve\_beaming\_[inst]} &  Positive semi-amplitude of the beaming effect, representing the term $A_1 \sin{\phi(t)}$, i.e. a modulation around the median flux level of the star & None \\
    \text{[comp]\_phase\_curve\_atmospheric\_[inst]} & Positive full (peak-to-peak) amplitude of the atmospheric contribution, representing the term $-2 B_1 (1 - \cos{\phi(t)}$, i.e. an additive component to the companion's nightside flux & None \\
    \text{[comp]\_phase\_curve\_atmospheric\_shift\_[inst]} & Time shift of the atmospheric contribution term (in days) & None \\
    \text{[comp]\_phase\_curve\_ellipsoidal\_[inst]} & Positive full (peak-to-peak) amplitude of the leading-order term of the ellipsoidal modulation, representing the term $-2 B_2 (1 - \cos{(2\phi(t))})$, i.e. an additive component to the system's flux from spherical (non-distorted) bodies  & None \\
    \text{[comp]\_phase\_curve\_ellipsoidal\_2nd\_[inst]} & Positive full (peak-to-peak) amplitude of the next-order term of the ellipsoidal modulation, representing the term $-2 B_3 (1 - \cos{(3\phi(t))})$, i.e. an additive component to the system's flux from spherical (non-distorted) bodies  & None \\
    
    &&\\
    \multicolumn{3}{c}{For differentiating thermal emission and reflected light one can use:} \\
    \text{[comp]\_phase\_curve\_atmospheric\_ thermal\_[inst]} & Positive full (peak-to-peak) amplitude of the atmospheric thermal emission & None \\
    \text{[comp]\_phase\_curve\_atmospheric\_ thermal\_shift\_[inst]} & Time shift of the atmospheric thermal emission (in days) & 0 \\
    \text{[comp]\_phase\_curve\_atmospheric\_ reflected\_[inst]} & Positive full (peak-to-peak) amplitude of the atmospheric reflected light & None \\
    \text{[comp]\_phase\_curve\_atmospheric\_reflected\_shift\_[inst]} & Time shift of the atmospheric reflected light (in days) & 0 \\
    
    &&\\
	\multicolumn{3}{r}{... continued on next page ...}
	\end{tabular}
\end{table*}

\addtocounter{table}{-1}
\begin{table*}[ht]
    \centering
	\caption{... continued from previous page ...}
	\centering
	\begin{tabular}{lp{10cm}l}
    \hline
    \hline
    Parameter & Explanation & Default\\    
    \hline
    &&\\
    \multicolumn{3}{c}{Star spots ($i=1,2,3...,$\text{N\_spots})} \\
    \hline
    \text{host\_spot\_[i]\_long\_[inst]} & Longitude of star spot number $i$ on the host (in degree from 0 to 360) & 0\\
    \text{host\_spot\_[i]\_lat\_[inst]} & Latitude of star spot number $i$ on the host (in degree from -90 to 90) & 0\\
    \text{host\_spot\_[i]\_size\_[inst]} & The angular radius of star spot number $i$ on the host (in degree) & 0\\
    \text{host\_spot\_[i]\_brightness\_[inst]} & The brightness ratio between star spot number $i$ and the surface of the host & 0\\

    \text{[comp]\_spot\_[i]\_long\_[inst]} & Longitude of star spot number $i$ on the companion (in degree from 0 to 360) & 0\\
    \text{[comp]\_spot\_[i]\_lat\_[inst]} & Latitude of star spot number $i$ on the companion (in degree from -90 to 90) & 0\\
    \text{[comp]\_spot\_[i]\_size\_[inst]} & The angular radius of star spot number $i$ on the companion (in degree) & 0\\
    \text{[comp]\_spot\_[i]\_brightness\_[inst]} & The brightness ratio between star spot number $i$ and the surface of the companion & 0\\
    
    &&\\
    \multicolumn{3}{c}{Stellar flares ($i=1,2,3...,$\text{N\_flares}) } \\
    \hline
    \text{flare\_tpeak\_[i]} & Peak time of flare number $i$ & 0\\
    \text{flare\_fwhm\_[i]} & Full-width at half maximum of flare number $i$ & 0\\
    \text{flare\_ampl\_[i]} & Amplitude of flare number $i$ & 0\\
    
    &&\\
    \multicolumn{3}{c}{Advanced parameters (for proficient users of \texttt{ellc})} \\
    \hline

    \text{[comp]\_q} & Mass ratio between the companion and host & 1\\
    
    \text{host\_gdc} & Gravity darkening coefficient for the host & None\\
    \text{[comp]\_gdc} & Gravity darkening coefficient for the companion & None\\
    
    \text{host\_atmo\_[inst]} & Coefficient of a simplified reflection and emission model on the host's side facing the companion & None\\
    \text{[comp]\_atmo\_[inst]} & Coefficient of a simplified reflection and emission model on the companion's side facing the host & None\\
    
    \text{host\_bfac\_[inst]} & Doppler boosting factor of the host & None\\
    \text{[comp]\_bfac\_[inst]} & Doppler boosting factor of the companion & None\\

    \text{didt\_[inst]} & Rate of change of inclination (in degrees per anomalistic period) & None\\
    \text{domdt\_[inst]} & Rate of apsidal motion (in degrees per anomalistic period) & None\\

    \text{host\_rotfac} & Asynchronous rotation factor for the host & None\\
    \text{[comp]\_rotfac} & Asynchronous rotation factor for the companion & None\\

    \text{host\_hf\_[inst]} & Fluid second Love number for radial displacement, for the host; only used if host\_shape\_[inst] is love & 1.5\\
    \text{[comp]\_hf\_[inst]} & Fluid second Love number for radial displacement, for the companion; only used if [comp]\_shape\_[inst] is love & 1.5\\

    \text{host\_lambda\_[inst]} & Sky-projected angle between orbital and rotation axes for the host (in degree) & None\\
    \text{[comp]\_lambda\_[inst]} & Sky-projected angle between orbital and rotation axes for the companion (in degree) & None\\

    \text{host\_vsini\_[inst]} & Rotational v sini for calculation of R-M effect for the host (in km/s) & None\\
    \text{[comp]\_vsini\_[inst]} & Rotational v sini for calculation of R-M effect for the companion (in km/s) & None\\
    
    &&\\
	\end{tabular}
    \begin{minipage}{8.7in} 
    \text{[comp]}: placeholder for the actual name given to the companion\\
    \text{[inst]}: placeholder for the actual name given to the instrument\\
    \text{[key]}: placeholder for either \textit{flux} or \textit{rv}\\
    \end{minipage}
\end{table*}

%% file: tab_Derived.tex
\clearpage
\begin{table*}[ht]
    \centering
	\caption{A list of all values that will be derived from the \texttt{allesfitter} posteriors. Note that not all of these can be derived every time. This list reflects \texttt{allesfitter} version 1.1.1. For future additions and the most up to date documentation see \url{www.allesfitter.com}.}
    \label{tab:Derived parameters}
	\centering
	\begin{tabular}{ll}
    \hline
    \hline
    Derived parameter & Equation \\
    \hline
    
    Host radius over semi-major axis; $R_\mathrm{host}/a_\mathrm{[comp]}$ & 
    r / (1 + k) \\
    
    Semi-major axis over host radius; $a_\mathrm{[comp]}/R_\mathrm{host}$ & 
    (1 + k) / r \\ 
    
    Companion radius over semi-major axis; $R_\mathrm{[comp]}/a_\mathrm{[comp]}$ & 
    r $\cdot$ k / (1 + k) \\
    
    Companion radius; $R_\mathrm{[comp]}$ ($\mathrm{R_{\oplus}}$) & 
    $R_h$ $\cdot$ k \\ 
    
    Companion radius; $R_\mathrm{[comp]}$ ($\mathrm{R_{jup}}$) & 
    $R_h$ $\cdot$ k \\ 
    
    Semi-major axis; $a_\mathrm{[comp]}$ ($\mathrm{R_{\odot}}$) & 
    $R_h$ / $R_h/a$ \\ 
    
    Semi-major axis; $a_\mathrm{[comp]}$ (AU) & 
    $R_h$ / $R_h/a$\\ 
    
    Inclination; $i_\mathrm{[comp]}$ (deg) & 
    $\arccos{( \cos{i} )}$ \\
    
    Eccentricity; $e_\mathrm{[comp]}$ & 
    $f_s^2 + f_c^2$ \\
    
    Argument of periastron; $w_\mathrm{[comp]}$ (deg) & 
    $\arctan2(f_s, f_c) \,\%\, 360^\circ$ \\
    
    Mass ratio; $q_\mathrm{[comp]}$ & 
    $\frac{1}{( a / a_1 ) - 1}$ with $a_1 = K_h \cdot P \cdot \frac{ \sqrt{(1 - e^2)} }{\sin(i)}$\\
     
    Companion mass; $M_\mathrm{[comp]}$ ($\mathrm{M_{\oplus}}$) &
    $q \cdot M_\mathrm{h}$ \\
    
    Companion mass; $M_\mathrm{[comp]}$ ($\mathrm{M_{jup}}$) & 
    $q \cdot M_\mathrm{h}$ \\
    
    Impact parameter; $b_\mathrm{tra;[comp]}$ & $\frac{a \cos i}{R_{h}}\left(\frac{1-e^{2}}{1+e \sin \omega}\right)$ \\
     
    Total transit duration (I to IV); $T_\mathrm{tot;[comp]}$ (h) & $\frac{P}{\pi} \sin ^{-1}\left[\frac{R_{h}}{a} \frac{\sqrt{(1+k)^{2}-b^{2}}}{\sin i}\right] \cdot \frac{\sqrt{1-e^{2}}}{1 + e \sin \omega}$\\
          
    Full-transit duration (II to III); $T_\mathrm{full;[comp]}$ (h) & 
    $\frac{P}{\pi} \sin ^{-1}\left[\frac{R_{*}}{a} \frac{\sqrt{(1-k)^{2}-b^{2}}}{\sin i}\right] \cdot \frac{\sqrt{1-e^{2}}}{1 - e \sin \omega}$\\
    
    Epoch of occultation; $T_\mathrm{occ; [comp]}$ (days) & 
    $\approx T_0 + \frac{P}{2}\left[1+\frac{4}{\pi} e \cos \omega\right]$ \\
    
    Impact parameter of occultation; $b_\mathrm{occ;[comp]}$ & 
    $\frac{a \cos i}{R_{*}}\left(\frac{1-e^{2}}{1-e \sin \omega}\right)$ \\
    
    Host density from orbit; $\rho_\mathrm{host;[comp]}$ (cgs) & 
    $\approx \frac{3 \pi}{G P^{2}}\left(\frac{a}{R_{*}}\right)^{3}$ if $k^3 < 0.01$ \\
    
    Companion surface gravity from orbit; $g_\mathrm{[comp]}$ (cgs) &
    $\frac{2 \pi}{P} \frac{\sqrt{1-e^{2}} K_h}{\left(R_\mathrm{[comp]} / a\right)^{2} \sin i}$\\
    
    Equilibrium temperature; $T_\mathrm{eq;[comp]}$ (K) & $T_\mathrm{eff; h} \cdot {\frac{(1-A)}{E}}^{1/4} \cdot \sqrt{ \frac{R_\mathrm{h}}{2 a}}$\\
     & with albedo $A=0.3$ and emissivity $E=1$\\ 
     
    Transit depth (dil.); $\delta_\mathrm{tr; dil; [comp]; [inst]}$ (ppt) & measured from model lightcurves \\ 
    
    Transit depth (undil.); $\delta_\mathrm{tr; undil; [comp]; [inst]}$ (ppt) & measured from model lightcurves \\ 
    
    Occultation depth (dil.); $\delta_\mathrm{occ; dil; [comp]; [inst]}$ (ppt) & measured from model lightcurves \\ 
    
    Occultation depth (undil.); $\delta_\mathrm{occ; undil; [comp]; [inst]}$ (ppt) & measured from model lightcurves \\ 
    
    Nightside flux (dil.); $F_\mathrm{nightside; dil; [comp]; [inst]}$ (ppt) & measured from model lightcurves \\ 
    
    Nightside flux (undil.); $F_\mathrm{nightside; undil; [comp]; [inst]}$ (ppt) & measured from model lightcurves \\ 
    
    Combined host density from all orbits; $\rho_\mathrm{host; combined}$ (cgs) & combination of all $\rho_\mathrm{host;[comp]}$ \\
    
    &\\
	\end{tabular}
    \begin{minipage}{\textwidth} 
    \text{[comp]}: placeholder for the actual name given to the companion\\
    \text{[inst]}: placeholder for the actual name given to the instrument\\
    \text{[key]}: placeholder for either \textit{flux} or \textit{rv}\\
    For readability, we define 
    $k$=\text{[comp]\_rr}, 
    $r$=\text{[comp]\_rsuma}, 
    $T_0$=\text{[comp]\_epoch}, 
    $P$=\text{[comp]\_epoch}, 
    $\cos{i}$=\text{[comp]\_cosi}, 
    $f_s$=\text{[comp]\_f\_s}, 
    $f_c$=\text{[comp]\_f\_c},
    and $D_0$=\text{dil\_[inst]}.
    Additionally, the [comp] suffixes in the equations were omitted (aside from $R_\mathrm{[comp]}$) and the host suffixes were replaced with h.
    For explanation of these equations see e.g. \citet{Winn2011} and references therein.
    Note that some values can only be derived if the host parameters are given in \textit{params\_star.csv}.\\
    \end{minipage}
\end{table*}